\documentclass[preprint,prd,aps,11pt,showpacs,nofootinbib,tightenlines,eqsecnum,aps,epsf,epsfig]{revtex4}
\usepackage{graphicx}

\begin{document}

\def\intdk{\int\frac{d^4k}{(2\pi)^4}}
\def\sla{\hspace{-0.17cm}\slash}

\graphicspath{{Wesszumino/}{Nonabelian/}{oneloop/}{feynmanrules/}}

\title{\Large{\bf Renormalization of Supersymmetric Field Theories\\ in Loop Regularization with String-mode Regulators}}
\author{Jian-Wei Cui, Yong Tang and Yue-Liang Wu}
\affiliation{Kavli Institute for Theoretical Physics China, Key
Laboratory of Frontiers in Theoretical Physics, Institute of
Theoretical Physics, Chinese Academy of Sciences, Beijing 100190,
China}

\begin{abstract}
By applying the recently developed Loop Regularization(LR) with
string-mode regulators to supersymmetric field theories, we
explicitly verify the supersymmetric Ward identities in several
supersymmetric models at one-loop level. It is interesting to
observe that supersymmetry is a so remarkable symmetry that the
supersymmetric Ward identities hold as long as a regularization
scheme is realized in the exact four dimensional space-time with
translational invariance for the momentum integration, and the gauge
symmetry can be maintained once the regularization scheme preserves
supersymmetry and satisfies the consistency condition for
logarithmic divergences. As a manifest demonstration, we carry out a
complete one-loop renormalization for the massive Wess-Zumino model
by adopting the LR method, it is found that all the quadratic
divergences cancel out and the relations among masses and coupling
constants hold after renormalization, which agrees with the
well-known non-renormalization theorem. It is concluded that the LR
method preserves not only gauge symmetry but also supersymmetry. A
simple and definite derivation of Majorana Feynman rules is found to
be very useful.
\end{abstract}

\maketitle

\section{Introduction}\label{intro}

Supersymmetry has attracted physicists for several decades since it
was proposed in 1970s\cite{first}. As it is well-known that symmetry
has played an important role in particle physics, and three of the
four basic forces in nature are governed by gauge symmetries and
have successfully been described by quantum field theory. While
quantum field theories are bothered by the infinities which must be
regularized to be well-defined. On the other hand, whether the
symmetries of classical Lagrangian still hold in the quantum level
remains an important issue, this is because some times it is
difficult to distinguish between a real anomaly and an apparent
violation of the symmetries due to the use of a symmetry-violating
regularization method. In general, when a symmetry of original
Lagrangian is still a symmetry of a full quantum effective action,
such a symmetry is regarded to be preserved in the quantum level,
but there are several exceptions such as chiral anomaly. Thus one
may ask whether supersymmetry is a symmetry of the full quantum
theory. This question has been studied in a
regularization-independent way in ref.\cite{algebraic renor}, and
the answer is yes. This means when investigating the quantum effects
of the supersymmetric theories, one must adopt a
supersymmetry-preserving regularization method.

Several regularization methods have been applied to supersymmetric
theories, such as dimensional reduction(DRED)\cite{DRED},
differential regularization\cite{DfR} and the so-called implicit
regularization\cite{IR}, among them DRED is the most common one. It
has been shown that DRED can preserve supersymmetry in several
models\cite{CJN, BHZ1996, HKS1999}. Strictly speaking, DRED is
mathematically inconsistent\cite{Siegel1980, stockinger2005} to be
applied to the supersymmetric theories, which is similar to the case
when it is applied to the chiral theories, this is because both
supersymmetry and the definition of $\gamma_5$ require an exact
dimension. A consistent regularization method that can be applied to
all possible cases in quantum field theories is needed. In this
sense, the recently developed loop regularization(LR) with
string-mode regulators\cite{LR1,LR2} may deserve a special
attention, it has successfully been applied to the calculations of
triangle anomaly of QED with clarifying the possible ambiguities
caused by $\gamma_5$\cite{YLMa}, the evaluation of a consistent
coefficient of the CPT and Lorentz symmetry breaking Chern-Simons
term\cite{YLMa2}, the computation of all the one-loop
renormalization constants for the non-Abelian gauge theory and the
determination for the coefficient of QCD $\beta$
function\cite{jwcui}, and the derivation of the chiral effective
field theory with a dynamically generated spontaneous symmetry
breaking\cite{YBDai}. The key concept of this new regularization
method is the introduction of the irreducible loop integrals(ILIs)
which are evaluated from Feynman integrals by using Feynman
parameter method.

It has been shown that the LR method can preserve the non-Abelian
gauge symmetry, and meanwhile maintain the divergent behavior of
original field theories. In particular, the LR method is realized in
the original four dimensional space-time with translational and
Lorentz invariance even if two intrinsic mass scales are introduced,
thus it can balance the bosonic and fermionic degrees automatically
and there is also no ambiguity about the definition of $\gamma_5$.
It is then believed that this method will preserve supersymmetry as
well. In this paper, we will investigate the applicability of LR
method in supersymmetric theories.

The paper is organized as follows: in section II, we briefly
introduce the symmetry-preserving loop regularization with
string-mode regulators. In sections \ref{masslessWZ} and
\ref{massiveWZ}, we will verify the supersymmetric Ward identities
for the massless Wess-Zumino model and massive Wess-Zumino
model\cite{wz} separately and show that the LR method indeed
respects the Ward identities. As Ward identity is the reflection of
symmetry in quantum level, we then arrive at the conclusion that the
LR method is also a supersymmetry-preserving regularization for
supersymmetric models. In section \ref{susygauge}, we consider the
super Yang-Mills theory as the testing ground to explicitly
demonstrate the supersymmetric Ward identity and show that the LR
method does preserve supersymmetry, and meanwhile the gauge symmetry
is maintained only requiring the consistency condition for
logarithmic divergences. Note that the conventional dimensional
regularization was shown to break the Ward identity in such a
model\cite{CJN}, thus an alternative check by using LR method in our
present paper is nontrivial. In particular, we will demonstrate that
as long as the Dirac algebra for $\gamma$ matrices are carried out
in four dimensional space-time, and the shift of integration
variable can be safely made, the supersymmetric Ward identities are
preserved, which is actually independent of any concrete
prescription of regularization methods. Namely, as long as the
regularization scheme is realized in four dimensional space-time
with translational invariance for momentum integrals, like the LR
method, it then preserves supersymmetry. In section \ref{case}, as
an explicit demonstration, we will carry out the one-loop
renormalization for the massive Wess-Zumino model by using the LR
method , and all the obtained results agree with the well-known
non-renormalization theorem. Our conclusions and remarks are given
in the last section. The detailed derivation of Majorana Feynman
rules is presented in the appendix.

\section{Symmetry-preserving Loop Regularization}

It has been shown in\cite{LR1,LR2} that all one loop Feynman
integrals can be evaluated into the following 1-fold ILIs by using
the Feynman parameterization method:
\begin{eqnarray}
I_{-2\alpha}&=&\intdk\frac{1}{(k^2-M^2)^{2+\alpha}},\nonumber\\
I_{-2\alpha\ \mu\nu}&=&\intdk\frac{k_{\mu}k_\nu}{(k^2-M^2)^{3+\alpha}},\hspace{8mm}\alpha=-1,0,1,2,... \nonumber \\
I_{-2\alpha\
\mu\nu\rho\sigma}&=&\intdk\frac{k_{\mu}k_{\nu}k_{\rho}k_{\sigma}}{(k^2-M^2)^{4+\alpha}}
\end{eqnarray}
with $I_2$ and $I_0$ corresponding to the quadratic and logarithmic
divergent integrals. Where the effective mass factor $M^2$ is a
function of the external momenta $p_i$, the masses of particles
$m_i$ and the Feynman parameters.

In general, the loop momentum independent $M^2$ can be extended to
include linear term in $k$, which can be understood as a part of the
definition of the ILIs in the LR. The reason is as follows: Let
$M^2(k)$ has the following general form including linear term in $k$
$$ M^2(k) = M^2 + 2xk.p$$
with $x$ an arbitrary parameter. Then
$$ k^2 + M^2(k) = k^2 + 2xp.k + M^2 =(k+xp)^2 + M^2 -p^2 = (k+xp)^2 + M^2(p)= k^{'2} + M^2(p) $$
with $M^2(p) = M^2-p^2$ which becomes independent of $k$, and $k' =
k + xp$ via translational invariance. Again the only thing must be
paid attention is that one must follow the definition of ILIs to
cancel out the $k^2$ in the numerator before regularization.

When the regularized 1-fold ILIs satisfy the following consistency
conditions\cite{LR1,LR2}:
\begin{eqnarray}
& & I_{2\mu\nu}^R = \frac{1}{2} g_{\mu\nu}\ I_2^R, \quad
I_{2\mu\nu\rho\sigma }^R = \frac{1}{8} (g_{\mu\nu}g_{\rho\sigma} +
g_{\mu\rho}g_{\nu\sigma} +
g_{\mu\sigma}g_{\rho\nu})\ I_2^R  , \nonumber \\
& & I_{0\mu\nu}^R = \frac{1}{4} g_{\mu\nu} \ I_0^R, \quad
I_{0\mu\nu\rho\sigma }^R = \frac{1}{24} (g_{\mu\nu}g_{\rho\sigma} +
g_{\mu\rho}g_{\nu\sigma} + g_{\mu\sigma}g_{\rho\nu})\ I_0^R .
\end{eqnarray}
the resulting loop corrections are gauge invariant. Here the
superscript "R" denotes the regularized ILIs.

Note that the introduction on the concept of irreducible loop
integrals (ILIs) is crucial in the loop
regularization\cite{LR1,LR2}, where it has been shown that all
Feynman loop integrals can be evaluated to be expressed by the ILIs.
From the definition of ILIs, one of the important properties is that
there should be no $k^2$ in the numerator of loop integration, all
the ILIs can be classified into the scalar type ILIs with the
following loop integration

\[ \frac{1}{(k^2 - M^2)^{\alpha}} \]
and the tensor type ILIs with the following loop integration

\[ \frac{k_{\mu}k_{\nu} \cdots k_{\rho}}{(k^2 - M^2)^{\alpha}} \]

In evaluating the Feynman loop integrals into ILIs,  one should
always perform the Dirac algebra and Lorentz index-contraction
firstly to obtain the ILIs defined by the above "simplest" forms for
the one loop case (for two loop and higher loop case, see ref.[11]).
Therefore, for the integration

\[ g^{\mu\nu} \cdot k_{\mu} k_{\nu}/(k^2-M^2)^2 \]
which should not be written as

\[ g^{\mu\nu} \cdot I_{2\mu\nu}  \]
but it must be expressed as

\[ k^2/(k^2-M^2)^2 \]
then rewriting the $k^2$ in the numerator into $(k^2-M^2)+M^2$ so as
to cancel out the first term by the denominator. Thus the above
Feynman loop integration is regarded to be evaluated into the ILIs
and is given by the following form before regularization

\[  g^{\mu\nu} \cdot k_{\mu} k_{\nu}/(k^2-M^2)^2  = I_2 + M^2 * I_0 \]

From the above illustration, it is seen that in the spirit of
''irreducible loop integrals" (ILIs), the integration

\[  g^{\mu\nu} \cdot k_{\mu} k_{\nu}/(k^2-M^2)^2 \]
is not an ILI, one should not regularize such a loop integration in
the loop regularization method.

A simple regularization prescription for the ILIs was realized to
yield the above consistency conditions, its procedure is that:
Rotating to the four dimensional Euclidean space of momentum,
replacing the loop integrating variable $k^2$ and the loop
integrating measure $\int{d^4k}$ in the ILIs by the corresponding
regularized ones $[k^2]_l$ and $\int[d^4k]_l$:
\begin{eqnarray}
& & \quad k^2 \rightarrow [k^2]_l \equiv k^2+M^2_l\ ,
\nonumber \\
& & \int{d^4k} \rightarrow \int[d^4k]_l \equiv \lim_{N,
M_l^2}\sum_{l=0}^{N}c_l^N\int{d^4k}
\end{eqnarray}
where $M_l^2$ ($ l= 0,1,\ \cdots $) may be regarded as the regulator
masses for the ILIs. The regularized ILIs in the Euclidean
space-time are then given by:
\begin{eqnarray}
I_{-2\alpha}^R&=& i (-1)^{\alpha} \lim_{N,
M_l^2}\sum_{l=0}^{N}c_l^N\intdk\frac{1}{(k^2 + M^2 + M_l^2)^{2+\alpha}},\nonumber\\
I_{-2\alpha\ \mu\nu}^R&=& -i (-1)^{\alpha} \lim_{N,
M_l^2}\sum_{l=0}^{N}c_l^N\intdk\frac{k_{\mu}k_\nu}{(k^2+M^2+M_l^2)^{3+\alpha}},\hspace{8mm}\alpha=-1,0,1,2,... \nonumber \\
I_{-2\alpha\ \mu\nu\rho\sigma}^R&=& i (-1)^{\alpha} \lim_{N,
M_l^2}\sum_{l=0}^{N}c_l^N\intdk\frac{k_{\mu}k_{\nu}k_{\rho}k_{\sigma}}{(k^2+M^2+M_l^2)^{4+\alpha}}
\end{eqnarray}
where the coefficients $c_l^N$ are chosen to satisfy the following
conditions:
\begin{eqnarray}
\lim_{N, M_l^2}\sum_{l=0}^{N}c_l^N(M_l^2)^n = 0 \quad
       (n= 0, 1, \cdots)\label{cl conditions}
\end{eqnarray}
with the notation $\lim_{N, M_l^2}$ denoting the limit $\lim_{N,
M_R^2\rightarrow \infty}$. One may take the initial conditions
$M_0^2 = \mu_s^2 = 0$ and $c_0^N = 1$ to recover the original
integrals in the limit $M_l^2 \to \infty$ ($l=1,2,\cdots$ ). Such a
new regularization is called as Loop Regularization (LR)
\cite{LR1,LR2}. The prescription in LR method is very similar to
Pauli-Villars prescription, but two concepts are totally different
as the prescription in the loop regularization is acting on the ILIs
rather than on the propagators in Pauli-Villars scheme. This is why
the Pauli-Villars regularization violates non-Abelian gauge
symmetry, while LR method can preserve non-Abelian gauge symmetry.

As the simplest solution of eq. (\ref{cl conditions}), taking the
string-mode regulators
\begin{equation}
M_l^2=\mu_s^2+lM_R^2
\end{equation}
with $l=1,2,\cdots$, the coefficients $c_l^N$ are completely
determined
\begin{equation}
 c_l^N=(-1)^l\frac{N!}{(N-l)!l!}\label{mus}
\end{equation}
Here $M_R$ may be regarded as a basic mass scale of loop regulator .
It has been shown in \cite{LR2} that the above regularization
prescription can be understood in terms of Schwinger proper time
formulation with an appropriate regulating distribution function.

With the string-mode regulators for $M_l^2$ and $c_l^N$ in above
equations, the regularized ILIs $I_2^R$ and $I_0^R$ can be evaluated
to the following explicit forms\cite{LR1,LR2}:
\begin{eqnarray}
I_2^R&=&\frac{-i}{16\pi^2}\{M_c^2-\mu^2[ln\frac{M_c^2}{\mu^2}-\gamma_w+1+y_2(\frac{\mu^2}{M_c^2})]\}  \nonumber \\
I_0^R&=&\frac{i}{16\pi^2}[ln\frac{M_c^2}{\mu^2}-\gamma_w+y_0(\frac{\mu^2}{M_c^2})]
\end{eqnarray}
with $\mu^2=\mu_s^2+M^2$, and
\begin{eqnarray}
& & \gamma_w \equiv \lim_{N}\{ \ \sum_{l=1}^{N} c_l^N \ln l +
     \ln [\ \sum_{l=1}^{N} c_l^N\ l \ln l \ ] \} = \gamma_E=0.5772\cdots, \nonumber \\
& & y_0(x)=\int_0^x d\sigma \frac{1-e^{-\sigma}}{\sigma},
 \quad  y_1(x)=\frac{e^{-x}-1+x}{x}\nonumber \\
& & y_2(x)=y_0(x)-y_1(x),\quad \lim_{x\rightarrow0}y_i(x)\rightarrow
0,\ i=0,1,2 \\
& & M_c^2\equiv \lim_{N,M_R} M_R^2 \sum_{l=1}^{N}c_l^N(l \ln l)
=\lim_{N,M_R}M_R^2/\ln N \nonumber
\end{eqnarray}
which indicates that the $\mu_s $ sets an IR `cutoff' at $M^2 =0$
and $M_c$ provides an UV `cutoff'. For renormalizable quantum field
theories, $M_c$ can be taken to be infinity
$(M_c\rightarrow\infty)$. In a theory without infrared divergence,
$\mu_s$ can safely run to $\mu_s=0$. Actually, in the case that
$M_c\to\infty$ and $\mu_s=0$, one recovers the initial integral.
Also once $M_R$ and $N$ are taken to be infinity, the regularized
theory becomes independent of the regularization prescription.  Note
that to evaluate the ILIs, the algebraic computing for multi
$\gamma$ matrices involving loop momentum $k\sla$ such as
$k\sla\gamma_{\mu}k\sla$ should be carried out to be expressed in
terms of the independent components: $\gamma_\mu$,
$\sigma_{\mu\nu}$, $\gamma_5\gamma_{\mu}$, $\gamma_5$.

We shall directly show that loop regularization is manifestly
translational invariant in spite of the existence of two energy
scales, which is a very important feature in applying to
supersymmetric theories in this paper. To see that, we shall verify
that the regularized ILIs should arrive at the same results whether
the loop regularization prescription is applied before or after
shifting the integration variables for momentum. For an explicit
illustration, let us examine a simple logarithmic divergent Feynman
integral:
\begin{eqnarray}
L={\intdk}\frac{1}{k^2-m_1^2}\frac{1}{(k-p)^2-m_2^2}
\end{eqnarray}
As the first step of loop regularization, we shall apply the general
Feynman parameter formula
\begin{eqnarray}
\frac{1}{a_1^{\alpha_1}a_2^{\alpha_2}{\cdots}a_n^{\alpha_n}}& = &
\frac{\Gamma(\alpha_1+\cdots+\alpha_n)}{\Gamma(\alpha_1)\cdots\Gamma(\alpha_n)}
\int_0^1dx_1\int_0^{x_1}dx_2\cdots\int_0^{x_{n-2}}dx_{n-1} \nonumber
\\
& &
\frac{(1-x_1)^{\alpha_1-1}(x_1-x_2)^{\alpha_2-1}{\cdots}x_{n-1}^{\alpha_n-1}}
{[a_1(1-x_1)+a_2(x_1-x_2)+\cdots+a_nx_{n-1}]^{\alpha_1+\cdots+\alpha_n}}
\end{eqnarray}
to the Feyman integral and obtain the following integral
\begin{eqnarray}
L&=&{\intdk}\int_0^1dx\frac{1}{\{(1-x)(k^2-m_1^2)+x[(k-p)^2-m_2^2]\}^2}\nonumber\\
&=&{\intdk}\int_0^1dx\frac{1}{\{(k-xp)^2-[(1-x)m_1^2+xm_2^2-x(1-x)p^2]\}^2}\nonumber\\
&=&\int_0^1dx{\intdk}\frac{1}{[ (k-xp)^2 -M^2]^2}
\end{eqnarray}
with $M^2=(1-x)m_1^2+xm_2^2-x(1-x)p^2$.

By making Wick rotation and applying the loop regularization
prescription before shifting the integration variable, i.e.,
rewriting the momentum factor $(k-xp)^2$ into $(k-xp)^2 = k^2 -
2xp.k + x^2p^2$, then replacing $k^2$ by $k^2 + M_l^2$, namely
\begin{equation}
(k-xp)^2 = k^2 - 2xp.k + x^2p^2 \to k^2 + M_l^2 - 2xp.k + x^2p^2 =
(k-xp)^2 + M_l^2
\end{equation}
we then obtain the regularized Feynman integral
\begin{eqnarray}
L^{R}= i \lim_{N, M_l^2}\sum_{l=0}^{N}c_l^N \int_0^1dx
\intdk\frac{1}{[(k-xp)^2+M^2+M_l^2]^2}
\end{eqnarray}
which becomes a well defined integral, so that we can safely shift
the integration variable:
\begin{eqnarray}
L^{R}=\int_0^1dx\lim_{N,
M_l^2}\sum_{l=0}^{N}c_l^N\intdk\frac{1}{(k^2+M^2+M_l^2)^2}
\end{eqnarray}
The same result can be arrived by using the standard procedure of
loop regularization with first shifting the integration variable for
momentum, which yields the standard scalar type ILI
\begin{eqnarray}
L_0&=& \int_0^1dx{\intdk}\frac{1}{( k^2 -M^2)^2} = \int_0^1dx\  I_0
\end{eqnarray}
after applying the loop regularization prescription, the same form
is reached
\begin{eqnarray}
L_0^R= i \int_0^1dx\lim_{N,
M_l^2}\sum_{l=0}^{N}c_l^N\intdk\frac{1}{(k^2+M^2+M_l^2)^2} \equiv
L^R
\end{eqnarray}
which shown that in loop regularization method, one can safely shift
the integration variables and express all the Feynman integrals in
terms of ILIs before applying for the regularization prescription.

From the above explicit demonstration, it is seen that the loop
regularization is indeed translational invariant. In fact, this
property also allows us to eliminate the ambiguities and make a
consistent calculation for the chiral anomaly even in the existence
of linear divergent integral\cite{YLMa,YLMa2}. The similar
verification of translational invariance can be extended to the
linearly and quadratically divergent integrals, which is presented
in the Appendix A.

The above proof can in generally be extended to higher loops based
on several theorems proved in ref.\cite{LR1}, especially based on
the theorem I, theorem V and theorem VI over there. The theorem I is
the so-called factorization theorem for overlapping divergences
which states that overlapping divergences which contain divergences
of sub-integrals and overall divergences in the general Feynman loop
integrals become completely factorizable in the corresponding ILIs.
The theorem V is the so-called reduction theorem for overlapping
tensor type integrals which states that the general overlapping
tensor type Feynman integrals of arbitrary loop graphs are
eventually characterized by the overall one-fold tensor type ILIs of
the corresponding loop graphs. This theorem is the key theorem for
the generalization of treatments and also for the prescriptions from
one loop graphs to arbitrary loop graphs. The theorem VI which is
the so-called relation theorem for tensor and scalar type ILIs which
states that for any fold tensor and scalar type ILIs, as long as
their power counting dimension of the integrating loop momentum are
the same, then the relations between the tensor and scalar type ILIs
are also the same and independent of the fold number of ILIs. This
theorem is crucial to extend the consistency conditions of gauge
invariance from divergent one loop ILIs to higher loop ILIs.

\section{Ward identity in Massless Wess-Zumino model}\label{masslessWZ}

\begin{figure}
   \includegraphics[scale=0.6,angle=0]{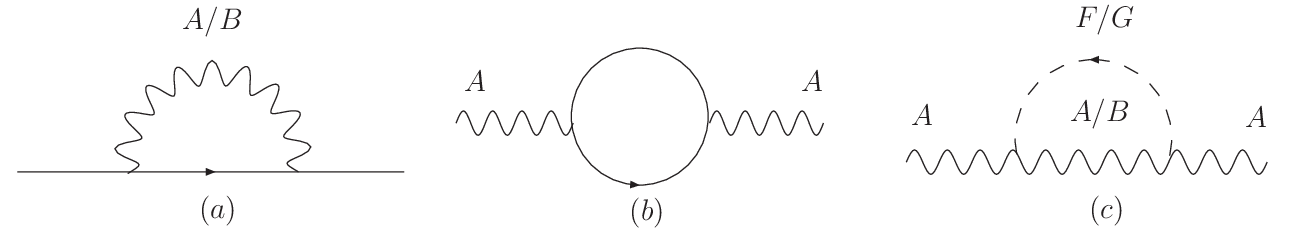}
    \caption{Three diagrams contribute at one-loop level}\label{threeinone1}
\end{figure}

We begin with the massless Wess-Zumino theory which is the simplest
supersymmetric model. The Lagrangian is:
\begin{eqnarray}\label{msls}
   L &=&-\frac{1}{2}\left( \partial _{\mu }A\right) ^{2}-\frac{1}{2}%
   \left( \partial _{\mu }B\right) ^{2}-\frac{1}{2}\overline{\chi}\partial\sla
   \chi +\frac{1}{2}F^{2}+\frac{1}{2}G^{2} \\
   &&+g\left[ -F\left( A^{2}-B^{2}\right) +2GAB+\overline{\chi}(A+i\gamma _{5}B)\chi \right]
\end{eqnarray}
the action is invariant, up to a total derivative, under the global
supersymmetric transformation shown below:
\begin{eqnarray}\label{transf}
   \delta A &=&\overline{\epsilon }\chi ,\;\delta B=-i\overline{\epsilon }
   \gamma _{5}\chi ,
   \nonumber \\
   \delta \chi &=&-\overline{\epsilon }\partial\sla (A+i\gamma _{5}B)
   +\overline{\epsilon }(F+i\gamma _{5}G),
   \nonumber \\
   \delta F &=&\overline{\epsilon }\partial\sla \chi ,\;\delta G=-i\overline{%
   \epsilon}\gamma _{5}\partial\sla \chi .
\end{eqnarray}
Using functional technique, one can deduce that the one-particle
irreducible(1PI) Green functions generating functional $\Gamma$ is
invariant under the supersymmetric transformation\cite{wzrnm}. The
supersymmetric Ward identity we choose to check is involving
two-point irreducible functions:
\begin{equation}\label{mlsdnt}
   \frac{\delta^{2}\Gamma}{\delta A(x)\delta A(y)}\delta_{\gamma\alpha}
   -\left(\partial\sla_{y}\right)_{\gamma\beta}\frac{\delta^{2}\Gamma }{%
   \delta \chi_{\alpha }\left( x\right)\delta \overline{\chi}_{\beta}\left(y\right)}=0
\end{equation}
This could be obtained from differentiating the equantion
$\delta\Gamma=0$ by $A(x)$ and $\bar{\chi}(x)$\cite{mless}. In the
momentum space, we can write it as:
 \begin{equation}\label{mlswi}
  \Gamma _{AA}(p)\delta _{\gamma \alpha }-i(p\sla)_{\gamma \beta }\Gamma
  _{\chi _{\alpha }\overline{\chi }_{\beta }}(p)=0
 \end{equation}
at one-loop level, Feynman diagrams contribute to this identity are
shown in FIG.\ref{threeinone1}.
\begin{eqnarray}
   \Gamma_{\chi_{\alpha}\overline{\chi}_{\beta}}^{(a)}(p) & = & 2\times4g^{2}\int
   \frac{d^{4}k}{(2\pi)^{4}}\frac{-\gamma_{\mu}k^{\mu}}{k^{2}}\frac{-i}{(k-p)^{2}}\nonumber \\
 & = &
 -i8g^{2}\int_{0}^{1}dx\int\frac{d^{4}l}{(2\pi)^{4}}\frac{x\gamma_{\mu}p^{\mu}}{[l^{2}-x(x-1)p^{2}]^{2}}\nonumber\\
 & =
 &-i8g^{2}\int_{0}^{1}dx{x\gamma_{\mu}p^{\mu}}I_0(x(x-1)p^{2})
\end{eqnarray}
the factor 2 appears because the wave line could be $A$ or $B$. And
the factor 4 results from the fact that the fermion is a Majorana
particle. We could discern this result more clearly from the Majorana
Feynman rules given in the appendix. According to the Feynman rules
we should calculate $\langle\chi\chi\rangle$ firstly, and then
obtain the $\langle\chi\bar{\chi}\rangle$ from the relation below:
\begin{eqnarray}
\langle\chi_i\bar{\chi_j}\rangle=\langle\chi_i(C^{-1}\chi)^T_j\rangle=\langle\chi_i\chi_k\rangle(-C^{-1}_{kj})
\end{eqnarray}

The calculation of $\Gamma_{AA}$ is straight forward:
\begin{eqnarray}
   \Gamma_{AA}^{(b)}(p) & = & -2g^{2}\int\frac{d^{4}k}{(2\pi)^{4}}\textrm{Tr}\left[\frac{\gamma_{\mu}k^{\mu}}{k^{2}}
   \frac{\gamma_{\upsilon}(k^{\nu}-p^{\nu})}{(k-p)^{2}}\right]\nonumber \\
& = &
8g^{2}\int\frac{d^{4}l}{(2\pi)^{4}}\left(\int_{0}^{1}dx\frac{xp^{2}}{[l^{2}-x(x-1)p^{2}]^{2}}-\frac{1}{l^{2}}\right)\nonumber\\
& = & 8g^{2}\left(\int_0^1dx{xp^{2}}I_0(x(x-1)p^{2})-I_2(0)\right)\\
   \Gamma_{AA}^{(c)}(p)&=&2\times4g^{2}\int\frac{d^{4}l}{(2\pi)^{4}}\frac{1}{l^{2}}=8g^2I_2(0)\label{qd0}
\end{eqnarray}

We can see immediately that the Ward identity(\ref{mlswi}) is
satisfied because the integrands cancel out. To arrive at above
results we have only carried out Dirac algebra for $\gamma$ matrices
in the four dimensional space-time and make the shift of the
integration variables. As these operations are all rational in a
four dimensional well-defined loop regularization method, thus we
conclude that at one-loop level the LR method indeed preserves
supersymmetric Ward identity in this simple model.

\section{Ward Identity in Massive Wess-Zumino model}\label{massiveWZ}

We are examining another supersymmetric model. The procedure is
similar to what we have done in the above massless model. The
Lagrangian of massive Wess-Zumino model is:
\begin{eqnarray}
  \nonumber
  L&=&-\frac{1}{2}\left(\partial_{\mu}A\right)^{2}-\frac{1}{2}
   \left(\partial_{\mu}B\right)^{2}-\frac{1}{2}\overline{\chi}\partial\sla\chi+\frac{1}{2}F^{2}+
   \frac{1}{2}G^{2}+m(AF-BG-\frac{1}{2}\overline{\chi}\chi)\\
   &&+g\left[-F\left(A^{2}-B^{2}\right)+2GAB+\overline{\chi}(A+i\gamma_{5}B)\chi\right]
\end{eqnarray}

It is different from the massless case with the mass term
$m(AF-BG-\frac{1}{2}\overline{\chi}\chi)$. In this model bosons and
fermions have equal masses as demanded by supersymmetry. In section
\ref{case}, we will explicitly show that the radiative corrections
do not violate such an equality. The supersymmetric transformation
of component fields are the same as Eq.(\ref{transf}). Following the
same procedure, the two-point Ward identity
 of this model is extended to be\cite{msive}:
\begin{equation}\label{msvwi}
  \Gamma_{AA}(p)\delta_{\gamma \alpha }-i(p\sla)_{\gamma \beta }\Gamma_{\chi_{\alpha}
  \overline{\chi}_{\beta}}(p)+i(p\sla)_{\gamma\alpha}\Gamma_{AF}(p)=0
\end{equation}

\begin{figure}
   \includegraphics[scale=0.6,angle=0]{threeinone.eps}
   \caption{The same as massless case}\label{threeinone}
\vspace{0.8cm}
   \centering
   \includegraphics[scale=0.5,angle=0]{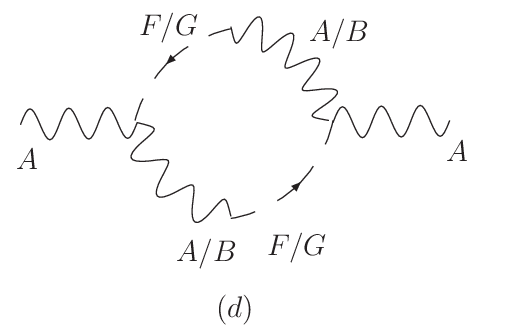}
   \caption{Additional diagram contributing to $\Gamma_{AA}$.}\label{boson2auxilary}
\vspace{0.8cm}
   \centering
   \includegraphics[scale=0.6,angle=0]{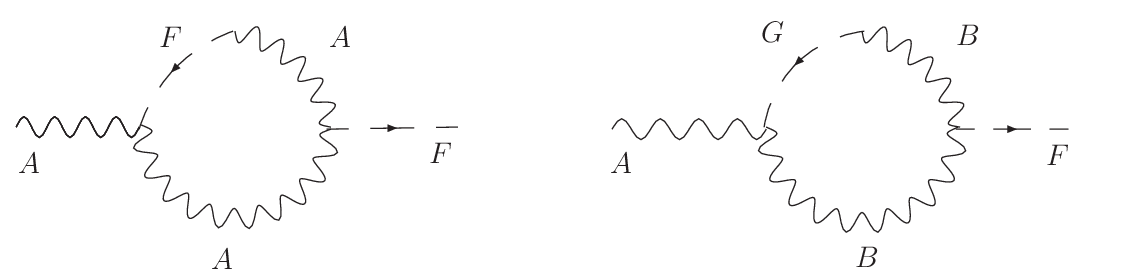}
   \caption{Two diagrams contributing to $\Gamma_{AF}$}\label{boson1auxilary}
\end{figure}

At one-loop level, the diagrams which contribute to this
supersymmetric Ward identity are shown in FIG.\ref{threeinone},
 FIG.\ref{boson2auxilary} and FIG.\ref{boson1auxilary}.

It is easy to show that two diagrams in Fig.(\ref{boson1auxilary})
contribute to $\Gamma_{AF}$ and their contributions cancel each
other:
\begin{eqnarray}
  \Gamma_{AF}=4g^{2}\int\frac{d^{4}k}{(2\pi)^{4}}\left[\frac{1}{k^{2}-m^{2}}\frac{m}{(k-p)^{2}-m^{2}}
  +\frac{1}{k^{2}-m^{2}}\frac{-m}{(k-p)^{2}-m^{2}}\right]=0
\end{eqnarray}

The calculations of other diagrams are straightforward, we just
present the final results as follows:
\begin{eqnarray}
\Gamma_{\chi_{\alpha}\overline{\chi}_{\beta}}^{(a)}(p) & = &
4g^{2}\int\frac{d^{4}k}{(2\pi)^{4}}
[\frac{-i\gamma_{\mu}k^{\mu}+m}{k^{2}-m^{2}}\frac{1}{(k-p)^{2}-m^{2}}
+(i\gamma_{5})\frac{-i\gamma_{\mu}k^{\mu}+m}{k^{2}-m^{2}}(i\gamma_{5})
\frac{1}{(k-p)^{2}-m^{2}}]\nonumber \\
 & = & 8g^{2}\int\frac{d^{4}l}{(2\pi)^{4}}\int_{0}^{1}dx\frac{-ix\gamma_{\mu}p^{\mu}}{[l^{2}-x(x-1)p^{2}-m^{2}]^{2}}\nonumber\\
 & = & 8g^2\int_0^1dx(-ix\gamma_{\mu}p^{\mu})I_0(x(x-1)p^2+m^2)\\
\Gamma_{AA}^{(b)}(p) & = &
-2g^{2}\int\frac{d^{4}k}{(2\pi)^{4}}\textrm{Tr}\left[\frac{-i\gamma_{\mu}k^{\mu}+m}{k^{2}-m^{2}}
\frac{-i\gamma_{\upsilon}(k^{\nu}-p^{\nu})+m}{(k-p)^{2}-m^{2}}\right]\nonumber \\
 & = & 8g^{2}\int\frac{d^{4}k}{(2\pi)^{4}}\int_{0}^{1}dx\left[\frac{1}{[l^{2}-x(x-1)p^{2}-m^{2}]}
 -\frac{2\left[m^{2}+x(1-x)p^{2}\right]}{[l^{2}-x(x-1)p^{2}-m^{2}]^{2}}\right]\nonumber\\
 & = &
 8g^2\int_o^1dx\left[I_2(x(x-1)p^2+m^2)-2\left[m^{2}+x(1-x)p^{2}\right]I_0(x(x-1)p^2+m^2)\right]\\
\Gamma_{AA}^{(c)}(p) & = & 2\times4g^{2}\int\frac{d^{4}k}{(2\pi)^{4}}\frac{-k^{2}}{k^{2}-m^{2}}\frac{1}{(k-p)^{2}-m^{2}}\nonumber \\
 & = & 8g^{2}\int\frac{d^{4}l}{(2\pi)^{4}}\int_{0}^{1}dx\left[\frac{-1}{[l^{2}-x(x-1)p^{2}-m^{2}]}
 +\frac{m^{2}+x(1-2x)p^{2}}{[l^{2}-x(x-1)p^{2}-m^{2}]^{2}}\right]\nonumber\\
 & = &
 8g^2\int_o^1dx\left[-I_2(x(x-1)p^2+m^2)+\left[m^{2}+x(1-2x)p^{2}\right]I_0(x(x-1)p^2+m^2)\right]\\
\Gamma_{AA}^{(d)}(p) & = & 2\times4g^{2}\int\frac{d^{4}k}{(2\pi)^{4}}\frac{m}{k^{2}-m^{2}}\frac{m}{(k-p)^{2}-m^{2}}\nonumber \\
 & = &
 8g^{2}\int\frac{d^{4}l}{(2\pi)^{4}}\int_{0}^{1}dx\frac{m^{2}}{[l^{2}-x(x-1)p^{2}-m^{2}]^{2}}\nonumber\\
 & = &8g^2\int_0^1m^2I_0(x(x-1)p^2+m^2)
\end{eqnarray}

Adding all the contributions together, we can see that the
integrands cancel out and the supersymmetric Ward identity holds.
Again, to arrive at above results we have only performed Dirac
algebra for $\gamma$ matrices in the four dimensional space-time and
make the shift of the integration variables. It further shows that
in the massive Wess-Zumino model the LR method can preserve
supersymmetry as well.

\section{Ward Identity in Supersymmetric gauge theory}\label{susygauge}

\begin{figure}
  \includegraphics[scale=0.6,angle=0]{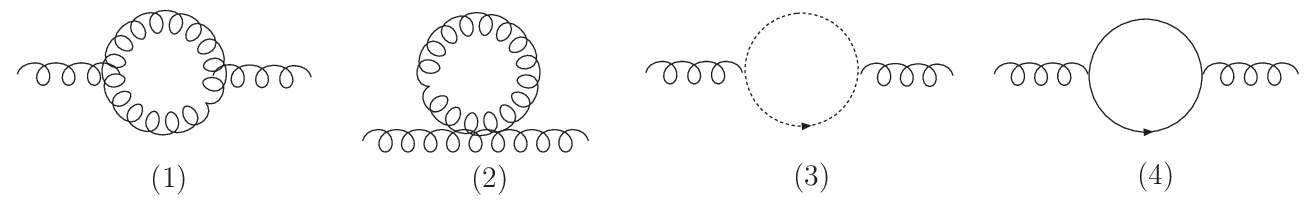}
  \caption{the 1st term}\label{twogauge}
\vspace{0.8cm}
  \includegraphics[scale=0.6]{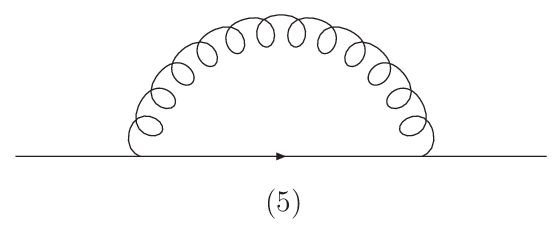}
  \includegraphics[scale=0.6]{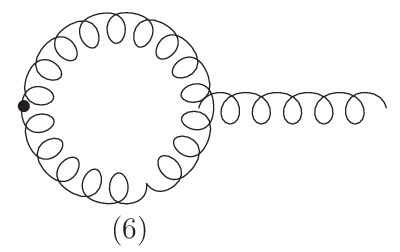}
  \caption{the 2nd term}\label{2ndline}
\vspace{0.8cm}
  \includegraphics[scale=0.6]{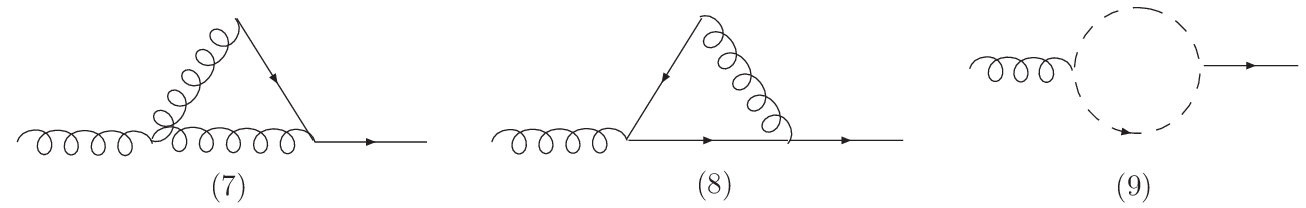}
  \caption{the 3rd and 4th terms}\label{gaugefermion}
\end{figure}

Let us consider a more complicated case, i.e., the supersymmetric
Yang-Mills theory. This model involves supersymmetry as well as
gauge symmetry. In the Wess-Zumino gauge, the Lagrangian (with
source terms) can be writen as:
\begin{equation}
     L=-\frac{1}{4}(F_{\mu\nu}^{a})^{2}-\frac{1}{2}(\partial^{\mu}A_{\mu}^a)+C^{\ast a}
     \partial D\sla^{ab}C^{b}-\frac{1}{2}\bar{\lambda}^{a} D\sla^{ab}\lambda^{b}
     +\frac{1}{2}D_{a}^{2}+J^{a\mu}A^{a}_{\mu}+\bar{J}^{a}\lambda^{a}+j_{D}^{a}D^{a}
\end{equation}
where $\lambda^a$ is a Majorana spinor and $D^a$ is the auxiliary
field. Similarly, the supersymmetric Ward identity is derived by
considering the functional variation of the Green function
generating functional under an infinitesimal supersymmetric
transformation. All the fields transform as follows:
\begin{eqnarray}
     \delta A^{a}_{\mu}&=& -\bar{\epsilon}\gamma_{\mu}\lambda^{a},\nonumber \\
     \delta\lambda^{a}&=&\sigma^{\mu\nu}F_{\mu\nu}^{a}\epsilon+i\gamma_{5}D^{a}\epsilon,\nonumber\\
     \delta D^{a}&=& \bar{\epsilon}i\gamma_{5} D\sla^{ab}\lambda^{b}
\end{eqnarray}
which lead to the following supersymmetric Ward
identity\cite{Nbel,CJN}:
\begin{eqnarray}\label{symwi}
            \nonumber
      0 &=& \frac{\delta J^{c'}_{\rho '}(z')}{\delta \hat{A}^{c}_{\rho}(z)}
            \frac{\delta \bar{J}^{b'}(y')}{\delta\hat{\bar{\lambda^{b}}}(y)}
            \langle \delta^{c'a}_{\rho '\mu}\delta^{4}(z'-x)(-\bar{\epsilon}\gamma_{\mu}\lambda^{a}(x))i\lambda^{b'}(y')
            \rangle\\
            \nonumber
         &&+\frac{\delta J^{c'}_{\rho '}(z')}{\delta \hat{A}^{c}_{\rho}(z)}
            \frac{\delta\bar{J}^{b'}(y')}{\delta\hat{\bar{\lambda^{b}}}(y)}
            \langle\delta^{b'a}\delta^{4}(y'-x)iA^{c'}_{\rho'}(z')\sigma^{\mu\nu}F^{a}_{\mu\nu}(x)\epsilon
            \rangle\\
            \nonumber
         &&+\frac{\delta J^{c'}_{\rho '}(z')}{\delta \hat{A}^{c}_{\rho}(z)}
            \frac{\delta\bar{J}^{b'}(y')}{\delta\hat{\bar{\lambda^{b}}}(y)}
            \langle\partial \cdot A^{a}(\texttt{x})\bar{\epsilon}\partial\sla\lambda^{a}(x)
            i\lambda^{b'}(y')iA_{\rho'}^{c'}(z')
            \rangle\\
            \nonumber
         &&+\frac{\delta J^{c'}_{\rho '}(z')}{\delta \hat{A}^{c}_{\rho}(z)}
            \frac{\delta\bar{J}^{b'}(y')}{\delta\hat{\bar{\lambda^{b}}}(y)}
            \langle i\lambda^{b'}(y')iA_{\rho'}^{c'}(z')
            (\partial_{\mu}C^{\ast a})f^{aef}\bar{\epsilon}\gamma_{\mu}\lambda^{e}(\texttt{x})C^{f}(x)
            \rangle\\
            \nonumber
         &&+\frac{\delta^{2}\bar{J}^{b'}(y')}{\delta\hat{A}^{c}_{\rho}(z)\delta\hat{\bar{\lambda^{b}}}(y)}
            \langle\partial \cdot A^{a}(\texttt{x})\bar{\epsilon}\partial\sla\lambda^{a}(x)
            i\lambda^{b'}(y')
            \rangle\\
            \nonumber
         &&+\frac{\delta^{2}\bar{J}^{b'}(y')}{\delta\hat{A}^{c}_{\rho}(z)\delta\hat{\bar{\lambda^{b}}}(y)}
            \langle i\lambda^{b'}(y')
            (\partial_{\mu}C^{\ast a})f^{aef}\bar{\epsilon}\gamma_{\mu}\lambda^{e}(\texttt{x})C^{f}(x)
            \rangle\\
\end{eqnarray}
here the notation $\langle...\rangle$ represents connected Green
functions and the integration over $x, y^\prime, z^\prime$ are
abbreviated. At the tree level, only the 1st, 2nd and 3rd terms in
the above equation contribute, one can easily verify that the
identity holds. At the one-loop level, only the 1st, 2nd, 3rd and
4th terms contribute, all the diagrams to this order are shown in
FIG.\ref{twogauge}, FIG.\ref{2ndline} and FIG.\ref{gaugefermion}.

We would like to point out that the first line of eq. (\ref{symwi})
is exactly the self-energy function of the gauge boson at one-loop
level, it can be seen from the relation below:
\begin{equation}
    \frac{\delta J^{c'}_{\rho'}(z')}{\delta
    \hat{A}^{c}_{\rho}(z)}=\Gamma_{A^{c'}_{\rho'}A^{c}_{\rho}}(z-z')
\end{equation}
Gauge symmetry requires this term to be transverse.

We now turn to the calculation of each term in the Ward identity,
and choose $\xi=1$ for simplicity.
\begin{eqnarray}
  \Pi_{\mu\nu}^{(1)} & = & -\frac{1}{2}g^{2}C_{acd}C_{bdc}\int\frac{d^{4}q}{\left(2\pi\right)^{4}}
  \frac{1}{q^{2}(q+p)^{2}}[10q_{\mu}q_{\nu}+5(p_{\mu}q_{\nu}+p_{\nu}q_{\mu})\nonumber \\
 &  & -2p_{\mu}p_{\nu}+(5p^{2}+2p\cdot q+2q^{2})g_{\mu\nu}]\nonumber \\
 & = & -\frac{1}{2}g^{2}C_{acd}C_{bdc}\int dx\int\frac{d^{4}l}{\left(2\pi\right)^{4}}
 \frac{\left[10l_{\mu}l_{\nu}+10x^{2}p_{\mu}p_{\nu}-7p_{\mu}p_{\nu}+4p^{2}g_{\mu\nu}+2(l^{2}+x^{2}p^{2})g_{\mu\nu}\right]}{[l^{2}-x(x-1)p^{2}]^{2}}\nonumber\\
 & = &-\frac{1}{2}g^{2}C_{acd}C_{bdc}\nonumber\\
 & &\int
 dx\left[(10I_{2\mu\nu}+2g_{\mu\nu}I_2)+(10x^{2}p_{\mu}p_{\nu}-7p_{\mu}p_{\nu}+4p^{2}g_{\mu\nu}+2x(2x-1)p^{2}g_{\mu\nu})I_0\right]\\
\Pi_{\mu\nu}^{(2)} & = &
g^{2}C_{acd}C_{bdc}\int\frac{d^{4}q}{\left(2\pi\right)^{4}}
\frac{3g_{\mu\nu}}{q^{2}}=g^{2}C_{acd}C_{bdc}\int
dx\int\frac{d^{4}l}{\left(2\pi\right)^{4}}
\frac{3g_{\mu\nu}(l^{2}+x^{2}p^{2})}{[l^{2}-x(x-1)p^{2}]^2}\nonumber\\
&=&g^{2}C_{acd}C_{bdc}\int dx\left[3g_{\mu\nu}I_2+3x(x-1)p^2g_{\mu\nu}I_0\right] \\
\Pi_{\mu\nu}^{(3)} & = &
g^{2}C_{acd}C_{bdc}\int\frac{d^{4}q}{\left(2\pi\right)^{4}}
\frac{p_{\mu}q_{\nu}+q_{\mu}q_{\nu}}{q^{2}(q+p)^{2}}\nonumber \\
 & = & g^{2}C_{acd}C_{bdc}\int dx\int\frac{d^{4}l}{\left(2\pi\right)^{4}}\frac{1}{[l^{2}-x(x-1)p^{2}]^{2}}
 [-\frac{1}{2}p_{\mu}p_{\nu}+l_{\mu}l_{\nu}+x^{2}p_{\mu}p_{\nu}]\nonumber\\
 &=&g^{2}C_{acd}C_{bdc}\int dx\left[I_{2\mu\nu}+(x^2-\frac{1}{2})p_\mu p_\nu
 I_0\right]\\
\Pi_{\mu\nu}^{(1+2+3)}&=&-g^{2}C_{acd}C_{bcd}\int
dx\left[(4x^{2}-3)(p^{2}g_{\mu\nu}-p_{\mu}p_{\nu})I_0-4I_{2\mu\nu}+2g_{\mu\nu}I_2\right]
\end{eqnarray}
Notice that in supersymmetric Yang-Mills theory, the fermions are
massless and belong to the adjoint representation of gauge group as
required by the fermion-boson symmetry. Then,
\begin{eqnarray}
  \nonumber
 \Pi_{\mu\nu}^{(4)} & = & -g^{2}4tr[T_{a}T_{b}]\int\frac{d^{4}q}{\left(2\pi\right)^{4}}
  \frac{(p+q)_{\mu}q_{\nu}+(p+q)_{\nu}q_{\mu}-(q^{2}+q\cdot p)g_{\mu\nu}}{q^{2}(q+p)^{2}}\\
 & = & g^{2}C_{acd}C_{bcd}\int dx\int\frac{d^{4}l}{\left(2\pi\right)^{4}}
 \left[\frac{(4x^{2}-2)(p^{2}g_{\mu\nu}-p_{\mu}p_{\nu})-4l_{\mu}l_{\nu}}{[l^{2}-x(x-1)p^{2}]^{2}}
 +\frac{2g_{\mu\nu}}{[l^{2}-x(x-1)p^{2}]}\right]\nonumber\\
 &=&g^{2}C_{acd}C_{bcd}\int
dx\left[(4x^{2}-2)(p^{2}g_{\mu\nu}-p_{\mu}p_{\nu})I_0-4I_{2\mu\nu}+2g_{\mu\nu}I_2\right]
\end{eqnarray}

Adding the four terms together, we obtain the self-energy of the gauge
boson which is gauge covariant:
\begin{eqnarray}
 \Pi_{\mu\nu}=g^{2}C_{acd}C_{bcd}(p^{2}g_{\mu\nu}-p_{\mu}p_{\nu})\int
 dxI_0
\end{eqnarray}
It is seen that the transverse condition of $\Pi_{\mu\nu}$ is
satisfied in supersymmetric model with the Feynman gauge $\xi =1$
gauge. The reason is that the quadratical divergences which will
potentially break the transverse condition cancel out in the
supersymmetric model. In fact, the cancelation of quadratical
divergences is a general feature of supersymmetric field theories,
it is also one of the motivations to propose supersymmetry. In other
word, if one wants to break supersymmetry but still maintain the
gauge symmetry, there are several ways to realize that, for
instance, give a mass to the fermion. In this case, the quadratical
divergences do not cancel out automatically and they may destroy the
transverse condition unless they can be regularized via an
appropriate regularization method to satisfy the consistency
conditions\cite{LR1}. As shown in \cite{LR1,LR2} the LR method is
competent in this case.

Note that here we have carried out the calculation in the Feynman
gauge with $\xi = 1$ for simplicity. In the general $\xi$ gauge,
there is a term which could break the transverse condition if the
regularization scheme does not satisfy the consistency condition for
the logarithmic divergences, the term is in proportion to
\[ (\xi - 1)*(a_0 - 1) \]
with $a_0$ being defined via logarithmic divergent $I_{0\mu\nu} =
\frac{1}{4} a_0 g_{\mu\nu} I_0$. In the Feynman gauge this term
vanishes due to $\xi = 1$. In the general $\xi$ gauge, it remains to
require the regularization scheme satisfy the consistency condition
for logarithmic divergent part, i.e., $a_0 =1$, so that the
transverse condition in gauge boson self-energy can hold.

And the fermion self-energy diagram is given by:
\begin{eqnarray}
  2\sigma^{\alpha\beta}p_{\beta}\Gamma^{(5)}_{\lambda\lambda}&=&-2\sigma^{\alpha\beta}p_{\beta}\int
  \frac{d^{4}q}{\left(2\pi\right)^{4}}C_{acd}C_{bcd}\times
  g^{2}\gamma^{\mu}\frac{i}{(q\sla+p\sla)}\gamma^{\nu}\frac{-i}{q^{2}}g_{\mu\nu}\nonumber\\
  &=&g^{2}C_{acd}C_{bcd}(p^{2}\gamma^{\alpha}-{p\sla} p^{\alpha})\int\frac{d^{4}l}{\left( 2\pi\right)^{4}}\int dx
  \frac{1}{[l^{2}-x(x-1)p^{2}]^{2}}\nonumber\\
  &=&g^{2}C_{acd}C_{bcd}(p^{2}\gamma^{\alpha}-{p\sla}
  p^{\alpha})\int dx I_0
\end{eqnarray}

There are two diagrams from the second term of Eq.(\ref{symwi}). The
non-linear part of $F_{\mu\nu}^{a}$ (Fig.\ref{2ndline}(6)) gives rise to the contribution:
\begin{eqnarray}
  {p\sla}\Pi_{A}^{(6)}(p)&=&{p\sla}g^{2}C_{acd}C_{bcd}\sigma^{\lambda\nu}\int\frac{d^{4}q}{\left(2\pi
  \right)^{4}}\frac{-ig_{\mu\nu}}{q^{2}}\frac{-ig_{\rho\lambda}}{(q+p)^{2}}
  [g^{\tau\rho}(p-q)^{\mu}+g^{\rho\mu}(2q+p)^{\tau}\nonumber\\
  &&-g^{\mu\tau}(2p+q)^{\rho}] \nonumber\\
  &=&\frac{3}{2}g^{2}C_{acd}C_{bcd}({p\sla} p^{\tau}-p^{2}\gamma^{\tau})\int\frac{d^{4}l}{\left(2\pi\right)^{4}}\int dx
  \frac{1}{[l^{2}-x(x-1)p^{2}]^{2}}\nonumber\\
  &=&\frac{3}{2}g^{2}C_{acd}C_{bcd}({p\sla} p^{\tau}-p^{2}\gamma^{\tau})\int
  dxI_0
\end{eqnarray}

To proceed, we consider the rest diagrams coming from the third and
fourth terms of Eq.(\ref{symwi}).
\begin{eqnarray}
 \Gamma^{(7)}_{\nu} & = & \int\frac{d^{4}q}{\left(2\pi\right)^{4}}[gC_{acd}\gamma^{\rho}]i(p\sla+q\sla)
  \frac{i}{(q\sla+p\sla)}[-i(p+q)^{\lambda}\frac{-ig_{\mu\lambda}}{(p+q)^{2}}]\nonumber \\
 &  & \times\lbrack g^{\nu k}(q-p)^{\mu}-g^{k\mu}(2q+p)^{\nu}+g^{\mu\nu}(2p+q)^{k}](-igC_{bdc})
 \frac{-ig_{\rho k}}{q^{2}}\nonumber \\
 & = & g^{2}C_{acd}C_{bcd}\int dx\int\frac{d^{4}l}{\left(2\pi\right)^{4}}\gamma_{\mu}
 \left[\frac{g^{\nu\mu}(l^{2}+x^{2}p^{2})-g^{\nu\mu}p^{2}-l^{\mu}l^{\nu}
 -x^{2}p^{\mu}p^{\nu}+p^{\mu}p^{\nu}}{[l^{2}-x(x-1)p^{2}]^{2}}\right]\nonumber\\
 &=&g^{2}C_{acd}C_{bcd}\int
 dx\gamma_{\mu}\left[g^{\mu\nu}I_2-I_2^{\mu\nu}+\left(x(2x-1)p^2-g^{\mu\nu}p^2+(1-x^2)p^{\mu}p^{\nu}\right)I_0\right]\\
\Gamma^{(8)}_{\nu} & = &
\int\frac{d^{4}q}{\left(2\pi\right)^{4}}[gC_{dac}\gamma^{\mu}]
\frac{i}{(q\sla+p\sla)}[gC_{bdc}\gamma^{\nu}]i\frac{iq\sla}{q\sla}\frac{-ig_{\mu\lambda}}{q^{2}}(iq^{\lambda})\nonumber \\
 & = &- g^{2}C_{acd}C_{bcd}\int dx\int\frac{d^{4}l}{\left(2\pi\right)^{4}}\gamma^{\nu}
 \left[\frac{l^{2}+x^{2}p^{2}-\frac{1}{2}p^{2}}{[l^{2}-x(x-1)p^{2}]^{2}}\right]\nonumber\\
 &=&- g^{2}C_{acd}C_{bcd}\int dx\gamma^{\nu}\left[I_2+\left(x(2x-1)p^2-\frac{1}{2}p^2\right)I_0\right]\\
\Gamma^{(9)}_{\nu} & = &
-g^{2}C_{acd}C_{bcd}\int\frac{d^{4}q}{\left(2\pi\right)^{4}}
\frac{iq^{\rho}\gamma_{\rho}}{q^{2}}i\left(q+p\right)^{\nu}\frac{1}{(q+p)^{2}}\nonumber \\
 & = & g^{2}C_{acd}C_{bcd}\int dx\int\frac{d^{4}q}{\left(2\pi\right)^{4}}\gamma_{\rho}
 \left[\frac{l^{\nu}l^{\rho}+x^{2}p^{\nu}p^{\rho}-\frac{1}{2}p^{\nu}p^{\rho}}{[l^{2}-x(x-1)p^{2}]^{2}}\right]\nonumber\\
 & = & g^{2}C_{acd}C_{bcd}\int dx\gamma_{\rho}\left[I_2^{\nu\rho}+\left(x^2-\frac{1}{2}\right)p^{\nu}p^{\rho}I_0\right]
\end{eqnarray}
the total contributions of three diagrams are found to be:
\begin{eqnarray}
-\frac{1}{2}g^{2}C_{acd}C_{bcd}(p^{2}\gamma^{\nu}-{p\sla}p^{\nu})\int
dxI_0
\end{eqnarray}

After taking into account of '$i$' factors from the formula, and
adding all the terms together, the integrands cancel out again,
which demonstrates that the supersymmetric Ward identity does hold.
To arrive this conclusion, we have only used the properties of four
dimensional $\gamma$ matrices and translational invariance of
momentum integrals. This implies that the LR method can indeed
preserve supersymmetry. The gauge symmetry holds only requiring the
consistency condition for logarithmic divergent part due to the
cancelation of quadratical divergences in supersymmetry-preserving
regularization method. In general, to preserve gauge symmetry in
non-supersymmetric models, it needs the consistency conditions for
both quadratic and logarithmic divergences for the regularized ILIs.
So far, we can conclude that the LR method preserves not only
non-Abelian gauge symmetry, but also supersymmetry.

\section{Renormalization of massive Wess-Zumino Model}\label{case}

In the previous sections we have shown that the LR method can
respect supersymmetric Ward identities in several models including
supersymmetric gauge theory, which implies that the LR method is
viable in supersymmetric theories. While in the above applications,
we have only used the main features of the LR method, namely the LR
method is realized in four dimensions with translational invariance
of momentum. In this section we shall apply the LR method to
manifestly perform one-loop renormalization for the massive
Wess-Zumino model. We choose such a model as a testing ground
because it is fairly simple and well-known. The model was shown to
be renormalizable to all orders in perturbation theory\cite{wzrnm}
by using higher derivative regularization. The same conclusion can
easily be obtained in the superspace formalism, where supergraph
Feynman rules of superfields greatly simplify the calculations. For
our purpose, we will use the component fields formalism to
renormalize the theory. This is because the superspace formalism
maintains supersymmetry in a manifest way, which is not suitable for
checking the consistency of a specific regularization scheme in
preserving supersymmetry. On the other hand, for the physically
interesting case of broken supersymmetry, it is usually preferred to
work with component fields.

The action of massive Wess-Zumino model is:
\begin{eqnarray}
S_{WZ}=\frac{1}{4}{\int}d^4xd^2\Theta(\frac{1}{8}\Phi\bar{D}\Phi-\frac{1}{2}m\Phi^2-\frac{1}{3}g\Phi^3)+h.c.\label{superlagr}
\end{eqnarray}
where $\Phi(x,\Theta,\bar{\Theta})$ is a chiral superfield. In terms
of component fields the Lagrangian can be written as:
\begin{eqnarray}
L&=&\frac{1}{2}(\partial{_\mu}A\partial^{\mu}A+\partial{_\mu}B\partial^{\mu}B+i\bar{\chi}\partial\sla\chi+F^2+G^2)-m(AF+BG+\frac{1}{2}\bar{\chi}\chi)\nonumber\\
&&-g[(A^2-B^2)F+2ABG+\bar{\chi}(A-i\gamma_5B)\chi]
\end{eqnarray}
The notions used here are slightly different from which in section
\ref{massiveWZ}. It is seen that the fields $F$ and $G$ have no
dynamical terms, they are auxiliary fields and can be integrated
out, which is equivalent to eliminate them from the Lagrangian by
using the equations of motions. In fact, in the building of
phenomenological supersymmetric model the auxiliary fields are
eliminated by
\begin{eqnarray}
F&=&mA+g(A^2-B^2)\\
G&=&mB+2gAB
\end{eqnarray}
Thus the Lagrangian can be written as:
\begin{eqnarray}
L&=&\frac{1}{2}(\partial_{\mu}A\partial^{\mu}A-m^2A^2)+\frac{1}{2}(\partial_{\mu}B\partial^{\mu}B-m^2B^2)+\frac{1}{2}\bar{\chi}(i\partial\sla-m)\chi\nonumber\\
&&-mgA(A^2+B^2)-g\bar{\chi}(A-i\gamma_5B)\chi-\frac{1}{2}g^2(A^2+B^2)^2\label{wzlagr}
\end{eqnarray}
which is the Lagrangian to be renormalized by using the LR method. The Lagrangian contains
one scalar particle $A$, one pesudoscalar particle $B$ and one
Majorana fermion $\chi$ with equal masses $m$.

Before proceeding, we will first check what supersymmetry can tell
us about the renormalization of massive Wess-Zumino model. The
answer can easily be yielded in the superfield formalism based on
the powerful supergraph technique. In the superfield formalism, the
non-renormalization theorem implies that up to any order of the
perturbative series only the first term (dynamical term) in
Eq(\ref{superlagr}) needs a counterterm due to the supersymmetry.
Namely, after renormalization the action gets the following form:
\begin{eqnarray}
S_{WZ}&=&\frac{1}{4}{\int}d^4xd^2\Theta(\frac{1}{8}\Phi\bar{D}\Phi-\frac{1}{2}m\Phi^2-\frac{1}{3}g\Phi^3+\frac{1}{8}\delta\Phi\bar{D}\Phi)+h.c.\nonumber\\
&=&\frac{1}{4}{\int}d^4xd^2\Theta(\frac{1}{8}Z\Phi\bar{D}\Phi-\frac{1}{2}\frac{m}{Z}Z\Phi^2-\frac{1}{3}\frac{g}{Z^{3/2}}Z^{3/2}\Phi^3)+h.c.
\end{eqnarray}
Where the $\delta$ term with $\delta=Z-1$ is a logarithmically
divergent counterterm, and $Z^{1/2}$ is the renormalization constant
of the superfield. In terms of component fields, the equations of
motion for $F$ and $G$ fields now become:
\begin{eqnarray}
F&=&\frac{1}{Z}[mA+g(A^2-B^2)]\\
G&=&\frac{1}{Z}(mB+2gAB)
\end{eqnarray}
After eleminating the auxiliary fields, it them leads to the renormalized Lagrangian:
\begin{eqnarray}
L&=&\frac{1}{2}Z(\partial_{\mu}A\partial^{\mu}A-(\frac{m}{Z})^2A^2)+\frac{1}{2}Z(\partial_{\mu}B\partial^{\mu}B-(\frac{m}{Z})^2B^2)+\frac{1}{2}Z\bar{\chi}(i\partial\sla-\frac{m}{Z})\chi\nonumber\\
&&-\frac{m}{Z}\frac{g}{Z^{3/2}}Z^{3/2}A(A^2+B^2)-\frac{g}{Z^{3/2}}Z^{3/2}\bar{\chi}(A-i\gamma_5B)\chi-\frac{1}{2}(\frac{g}{Z^{3/2}})^2Z^2(A^2+B^2)^2\label{wzlagr2}
\end{eqnarray}
Which shows that the renormalizations of fields, mass and coupling
constant must satisfy:
\begin{eqnarray}
\phi_{bare}=Z^{1/2}\phi; \hspace{1cm} m_{bare}=Z^{-1}m; \hspace{1cm}
g_{bare}=Z^{-3/2}g.\label{renorquani}
\end{eqnarray}
with $\phi=A, B, \chi$. We may summarize the features of the model:
i). This model is renormalizable, and after renormalization all the
vertexes are remained to be only one coupling constant. ii). The
fields, mass and coupling constant share a common renormalization
constant, which only contains logarithmical divergence. The
cancellation of quadratical divergence is a general feature of all
supersymmetric theories. iii). As required by supersymmetry, the
masses of bosons still equal to the mass of fermion after
renormalization.

\begin{figure}[ht]
\includegraphics[scale=0.7]{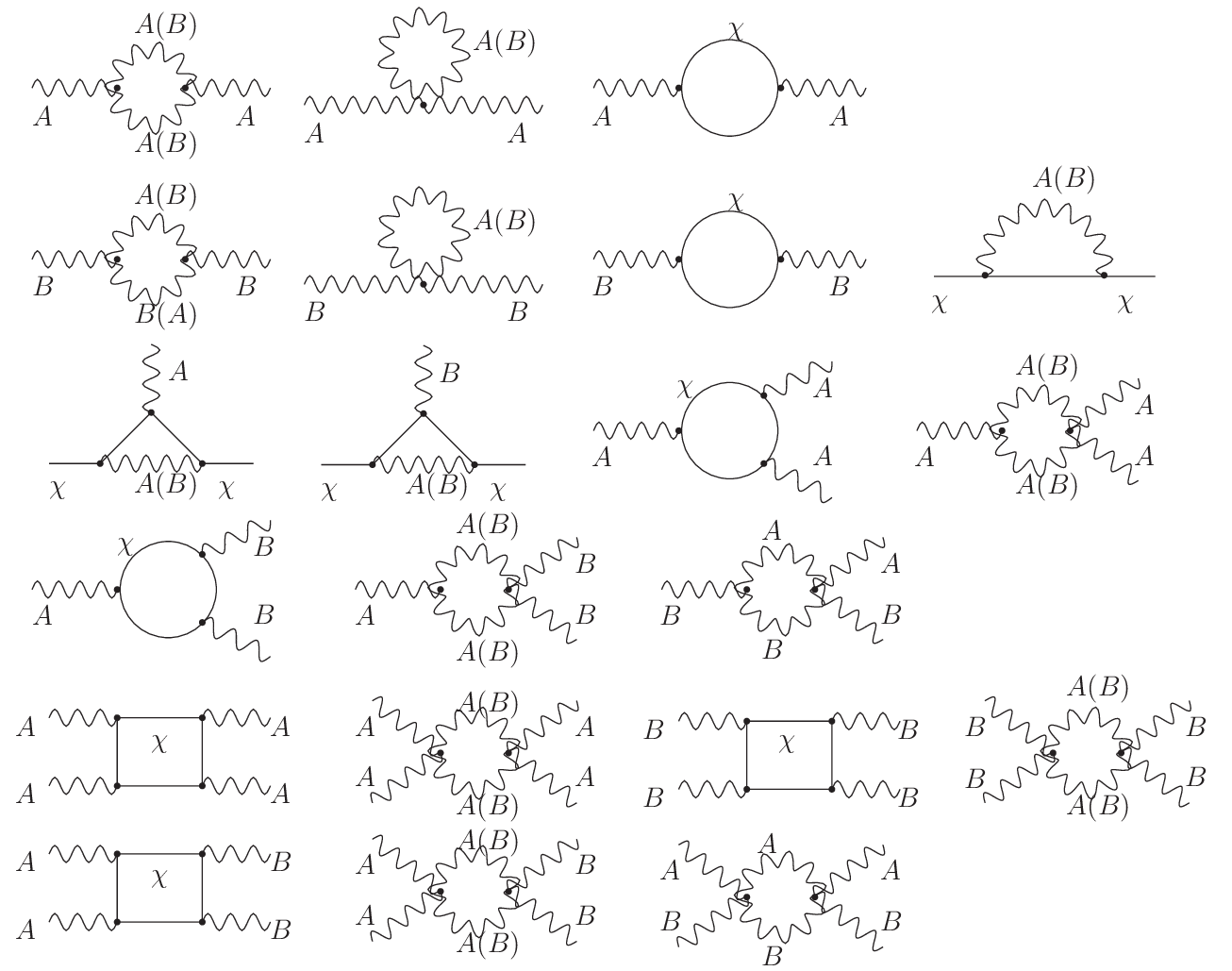}
\caption{non-vanishing one-loop divergent graphs in massive
Wess-Zumino model}\label{oneloop}
\end{figure}

Let us now make a detailed calculation for one-loop renormalization by using the LR Method.
The Feynman rules of Lagrangian(Eq.(\ref{wzlagr})) are listed in
the appendix, there are 7 types of vertices. What we are going to demonstrate is
that after renormalization all these 7 types of vertices
will get the same renormalized coupling constant, and all the
renormalization constants satisfy Eq.(\ref{renorquani}). It is easy
to verify that one-loop contributions to $\langle A \rangle$,
$\langle B \rangle$, $\langle AB \rangle$, $\langle AAB \rangle$,
$\langle BBB \rangle$, $\langle AAAB \rangle$, $\langle ABBB
\rangle$ are vanishing. The rest of divergent diagrams at one-loop
level are shown in FIG.\ref{oneloop}, the
permutation graphs are not presented for simplicity.

The field strength and mass renormalizations of filed $A$ can be
obtained from the calculations of two point Green function  $\langle
AA \rangle$. Five diagrams can contribute to $\langle AA \rangle$,
the total contribution is found to be:
\begin{eqnarray}
L_{\langle AA \rangle}&=&\frac{1}{2}\intdk[(-6img)^2+(-2img)^2]\frac{i}{k^2-m^2}\frac{1}{(k+p)^2-m^2}\nonumber\\
&&+\frac{1}{2}\intdk(-12ig^2-4ig^2)\frac{i}{k^2-m^2}\nonumber\\
&&-\frac{1}{2}tr\intdk(2igC^\dag)\frac{i}{k\sla-m}C^T(2igC^\dag)\frac{i}{(k\sla+p\sla)-m}C^T\nonumber\\
&=&4g^2\int_0^1\intdk\frac{2(1-x)p^2+m^2}{[k^2-m^2-x(x-1)p^2]^2}\nonumber\\
&=&4g^2\int_0^1\intdk[2(1-x)p^2+m^2]I_0
\end{eqnarray}
which is only logarithmic divergent as the quadratical divergences cancel out.
Using the loop regularization, the regularized $I_0$ has the following explicit form:
\begin{eqnarray}
I_0^R&=&\frac{i}{16\pi^2}[ln\frac{M_c^2}{\mu^2}-\gamma_\omega+y_0(\frac{\mu^2}{M_c^2})]
\end{eqnarray}
We shall adopt a subtraction scheme similar to the Minimal Subtraction
scheme in dimensional regularization. For that, it is useful to introduce an
arbitrary energy scale parameter $\mu_s$ and write $I_0^R$ as:
\begin{eqnarray}
I_0^R=\frac{i}{16\pi^2}ln\frac{M_c^2}{\mu_s^2}+\frac{i}{16\pi^2}[ln\frac{\mu_s^2}{\mu^2}-\gamma_\omega+y_0(\frac{\mu^2}{M_c^2})]\label{I0R}
\end{eqnarray}
then the divergent terms proportional to
$\frac{i}{16\pi^2}ln\frac{M_c^2}{\mu_s^2}$ for $M_c\longrightarrow\infty$ in the Feynman integral
are canceled by counterterms. As such a divergent term is independent of the Feynman
parameters $x$, we can integrate $x$ easily and obtain the divergent
part of these diagrams:
\begin{eqnarray}
L_{\langle AA \rangle;
div}=\frac{i}{4\pi^2}g^2(p^2+m^2)ln\frac{M_c^2}{\mu_s^2}
\end{eqnarray}
The counterterms correspondig to this divergence is:
\begin{eqnarray}
\delta L=\frac{1}{2}\delta_A(\partial_\mu A\partial^\mu
A)-\frac{1}{2}\delta_{m_A}m^2A^2
\end{eqnarray}
where
\begin{eqnarray}
\delta_A=-\frac{1}{4\pi^2}g^2ln\frac{M_c^2}{\mu_s^2}\label{z1};\hspace{0.5cm}
\delta_{m_A}=\frac{1}{4\pi^2}g^2ln\frac{M_c^2}{\mu_s^2}\label{z2}
\end{eqnarray}
from this we finally get:
\begin{eqnarray}
A_{bare}=(1-\frac{1}{8\pi^2}g^2ln\frac{M_c^2}{\mu_s^2})A=z^{1/2}A;\hspace{0.5cm}
m_{Abare}=(1+\frac{1}{4\pi^2}g^2ln\frac{M_c^2}{\mu_s^2})m=z^{-1}m\label{A}
\end{eqnarray}
where
\begin{eqnarray}
z=1-\frac{1}{4\pi^2}g^2ln\frac{M_c^2}{\mu_s^2}\label{z3}
\end{eqnarray}

The calculation for $\langle BB\rangle$ is similar, which gives:
\begin{eqnarray}
B_{bare}=(1-\frac{1}{8\pi^2}g^2ln\frac{M_c^2}{\mu_s^2})B=z^{1/2}B;\hspace{0.5cm}
m_{Bbare}=(1+\frac{1}{4\pi^2}g^2ln\frac{M_c^2}{\mu_s^2})m=z^{-1}m\label{B}
\end{eqnarray}

We now turn to the calculation of $\langle\chi\chi\rangle$, from the
Fig.(\ref{oneloop}) we can read directly:
\begin{eqnarray}
L_{\langle\chi\chi\rangle}&=&\intdk(2gC^{\dag}i)\frac{i}{k\sla-m}C^T(2gC^{\dag}i)\frac{i}{(k-p)^2-m^2}\nonumber\\
&&+\intdk(2gC^{\dag}\gamma_5)\frac{i}{k\sla-m}C^T(2gC^{\dag}\gamma_5)\frac{i}{(k-p)^2-m^2}\nonumber\\
&=&\int_0^1dx\intdk\frac{-8xg^2C^{\dag}p\sla}{(k^2-m^2-x(x-1)p^2)^2}\nonumber\\
&=&-\frac{i}{4\pi^2}g^2C^{\dag}{p\sla}ln\frac{M_c^2}{\mu_s^2}+...
\end{eqnarray}
We need the following counterterms to cancel this divergence:
\begin{eqnarray}
\delta
L=\frac{1}{2}\delta_\chi\bar{\chi}i\partial\sla\chi-\frac{1}{2}\delta_{m_\chi}m\bar{\chi}\chi
\end{eqnarray}
where:
\begin{eqnarray}
\delta_\chi=-\frac{1}{4\pi^2}g^2ln\frac{M_c^2}{\mu_s^2}\label{z4};\hspace{0.5cm}
\delta_{m\chi}=0
\end{eqnarray}
which indicates that the renormalization of field and mass is given
by:
\begin{eqnarray}
\chi_{bare}=(1-\frac{1}{8\pi^2}g^2ln\frac{M_c^2}{\mu_s^2})\chi=z^{1/2}\chi;\hspace{0.5cm}
m_{\chi
bare}=(1+\frac{1}{4\pi^2}g^2ln\frac{M_c^2}{\mu_s^2})m=z^{-1}m\label{chi}
\end{eqnarray}

So far, we have worked out the renormalization constants for the
fields $A$, $B$ and $\chi$ and their masses, the results agree with
Eq.(\ref{renorquani}). Let us switch to the renormalization of
coupling constant. As mentioned above, there are 7 types of vertices
which should be described by only one coupling constant when
supersymmetry holds. The contributions from all divergent diagrams
shown in FIG.\ref{oneloop} are found to be:
\begin{eqnarray}
L_{\langle AAA\rangle}&=&-tr\intdk(2gC^{\dag}i)\frac{i}{k\sla-m}C^T(2gC^{\dag}i)\frac{i}{(k\sla-p\sla_1)-m}C^T(2gC^{\dag}i)\frac{i}{(k\sla+p\sla_2)-m}C^T\nonumber\\
&&+\frac{1}{2}\intdk(-6img)\frac{i}{k^2-m^2}\frac{i}{(k+p_1)^2-m^2}(-12ig^2)+(p_1\rightarrow p_2)+(p_1\rightarrow p_3)\nonumber\\
&&+\frac{1}{2}\intdk(-2img)\frac{i}{k^2-m^2}\frac{i}{(k+p_1)^2-m^2}(-4ig^2)+(p_1\rightarrow p_2)+(p_1\rightarrow p_3)\nonumber\\
&=&i\frac{3}{2\pi^2}mg^3ln\frac{M_c^2}{\mu_s^2}+finite\ terms\\
L_{\langle ABB\rangle}&=&-tr\intdk(2gC^{\dag}i)\frac{i}{k\sla-m}C^T(2gC^{\dag}\gamma_5)\frac{i}{(k\sla-p\sla_1)-m}C^T(2gC^{\dag}\gamma_5)\frac{i}{(k\sla+p\sla_2)-m}C^T\nonumber\\
&&+\frac{1}{2}\intdk(-6img)\frac{i}{k^2-m^2}\frac{i}{(k+p_1)^2-m^2}(-4ig^2)\nonumber\\
&&+\frac{1}{2}\intdk(-2img)\frac{i}{k^2-m^2}\frac{i}{(k+p_1)^2-m^2}(-12ig^2)\nonumber\\
&&+\frac{1}{2}\intdk(-2img)\frac{i}{k^2-m^2}\frac{i}{(k+p_2)^2-m^2}(-4ig^2)+(p_2\rightarrow p_3)\nonumber\\
&=&i\frac{1}{2\pi^2}mg^3ln\frac{M_c^2}{\mu_s^2}+finite\ terms\\
L_{\langle A\chi\chi\rangle}&=&\intdk(2gC^{\dag}i)\frac{i}{(k\sla+p\sla_2)-m}C^T(2gC^{\dag}i)\frac{i}{(k\sla+p\sla_1)-m}C^T(2gC^{\dag}i)\frac{i}{k^2-m^2}\nonumber\\
&&+\intdk(2gC^{\dag}\gamma_5)\frac{i}{(k\sla+p\sla_2)-m}C^T(2gC^{\dag}i)\frac{i}{(k\sla+p\sla_1)-m}C^T(2gC^{\dag}\gamma_5)\frac{i}{k^2-m^2}\nonumber\\
&=&finite\ terms\\
L_{\langle B\chi\chi\rangle}&=&\intdk(2gC^{\dag}i)\frac{i}{(k\sla+p\sla_2)-m}C^T(2gC^{\dag}\gamma_5)\frac{i}{(k\sla+p\sla_1)-m}C^T(2gC^{\dag}i)\frac{i}{k^2-m^2}\nonumber\\
&&+\intdk(2gC^{\dag}\gamma_5)\frac{i}{(k\sla+p\sla_2)-m}C^T(2gC^{\dag}\gamma_5)\frac{i}{(k\sla+p\sla_1)-m}C^T(2gC^{\dag}\gamma_5)\frac{i}{k^2-m^2}\nonumber\\
&=&finite\ terms\\
L_{\langle AAAA\rangle}&=&-tr\intdk(2gC^{\dag}i)\frac{i}{k\sla-m}C^T(2gC^{\dag}i)\frac{i}{(k\sla+p\sla_2)-m}C^T(2gC^{\dag}i)\frac{i}{(k\sla+p\sla_2+p\sla_3)-m}\nonumber\\
&&C^T(2gC^{\dag}i)\frac{i}{(k\sla+p\sla_2+p\sla_3+p\sla_4)-m}C^T+(p_2\leftrightarrow p_3)+(p_3\leftrightarrow p_4)\nonumber\\
&&+\frac{1}{2}(-12ig^2)^2\frac{i}{k^2-m^2}\frac{i}{(k+p_1+p_2)^2-m^2}+(p_2\rightarrow p_3)+(p_2\rightarrow p_4)\nonumber\\
&&+\frac{1}{2}(-4ig^2)^2\frac{i}{k^2-m^2}\frac{i}{(k+p_1+p_2)^2-m^2}+(p_2\rightarrow p_3)+(p_2\rightarrow p_4)\nonumber\\
&=&i\frac{3}{\pi^2}g^4ln\frac{M_c^2}{\mu_s^2}+finite\ terms\\
L_{\langle BBBB\rangle}&=&-tr\intdk(2gC^{\dag}\gamma_5)\frac{i}{k\sla-m}C^T(2gC^{\dag}\gamma_5)\frac{i}{(k\sla+p\sla_2)-m}C^T(2gC^{\dag}\gamma_5)\frac{i}{(k\sla+p\sla_2+p\sla_3)-m}\nonumber\\
&&C^T(2gC^{\dag}\gamma_5)\frac{i}{(k\sla+p\sla_2+p\sla_3+p\sla_4)-m}C^T+(p_2\leftrightarrow p_3)+(p_3\leftrightarrow p_4)\nonumber\\
&&+\frac{1}{2}(-12ig^2)^2\frac{i}{k^2-m^2}\frac{i}{(k+p_1+p_2)^2-m^2}+(p_2\rightarrow p_3)+(p_2\rightarrow p_4)\nonumber\\
&&+\frac{1}{2}(-4ig^2)^2\frac{i}{k^2-m^2}\frac{i}{(k+p_1+p_2)^2-m^2}+(p_2\rightarrow p_3)+(p_2\rightarrow p_4)\nonumber\\
&=&i\frac{3}{\pi^2}g^4ln\frac{M_c^2}{\mu_s^2}+finite\ terms\\
L_{\langle AABB\rangle}&=&-tr\intdk(2gC^{\dag}i)\frac{i}{k\sla-m}C^T(2gC^{\dag}i)\frac{i}{(k\sla+p\sla_2)-m}C^T(2gC^{\dag}\gamma_5)\frac{i}{(k\sla+p\sla_2+p\sla_3)-m}\nonumber\\
&&C^T(2gC^{\dag}\gamma_5)\frac{i}{(k\sla+p\sla_2+p\sla_3+p\sla_4)-m}C^T\nonumber\\
&&-tr\intdk(2gC^{\dag}i)\frac{i}{k\sla-m}C^T(2gC^{\dag}i)\frac{i}{(k\sla+p\sla_2)-m}C^T(2gC^{\dag}\gamma_5)\frac{i}{(k\sla+p\sla_2+p\sla_4)-m}\nonumber\\
&&C^T(2gC^{\dag}\gamma_5)\frac{i}{(k\sla+p\sla_2+p\sla_3+p\sla_4)-m}C^T\nonumber\\
&&-tr\intdk(2gC^{\dag}i)\frac{i}{k\sla-m}C^T(2gC^{\dag}\gamma_5)\frac{i}{(k\sla+p\sla_3)-m}C^T(2gC^{\dag}i)\frac{i}{(k\sla+p\sla_3+p\sla_2)-m}\nonumber\\
&&C^T(2gC^{\dag}\gamma_5)\frac{i}{(k\sla+p\sla_2+p\sla_3+p\sla_4)-m}C^T\nonumber\\
&&+\frac{1}{2}(-4ig^2)(-12ig^2)\frac{i}{k^2-m^2}\frac{i}{(k+p_1+p_2)^2-m^2}\nonumber\\
&&+\frac{1}{2}(-12ig^2)(-4ig^2)\frac{i}{k^2-m^2}\frac{i}{(k+p_1+p_2)^2-m^2}\nonumber\\
&&+(-4ig^2)(-4ig^2)^2\frac{i}{k^2-m^2}\frac{i}{(k+p_1+p_3)^2-m^2}+(p_3\rightarrow p_4)\nonumber\\
&=&i\frac{1}{\pi^2}g^4ln\frac{M_c^2}{\mu_s^2}+finite\ terms\\
\end{eqnarray}
Introducing the following counterterms:
\begin{eqnarray}
\delta
L&=&-\delta_1mgA^3-\delta_2mgAB^2-\delta_3gA\bar{\chi}\chi-\delta_4gB\bar{\chi}i\gamma_5\chi\nonumber\\
& &-\delta_5\frac{1}{2}g^2A^4-\delta_6\frac{1}{2}g^2B^4-\delta_7g^2A^2B^2
\end{eqnarray}
with:
\begin{eqnarray}
\delta_1&=&\delta_2=\frac{1}{4\pi^2}g^2ln\frac{M_c^2}{\mu_s^2}\nonumber\\
\delta_3&=&\delta_4=0\nonumber\\
\delta_5&=&\delta_6=\delta_7=\frac{1}{4\pi^2}g^2ln\frac{M_c^2}{\mu_s^2}
\end{eqnarray}
It is easy to check that all the renormalized vertices lead to a single renormalization constant:
\begin{eqnarray}
g_{bare}=(1+\frac{3}{8\pi^2}g^2ln\frac{M_c^2}{\mu_s^2})g=z^{-3/2}g\label{g}
\end{eqnarray}
This equation, together with Eq.(\ref{A}), Eq.(\ref{B}) and
Eq.(\ref{chi}), shows that the LR method works well in the perturbative theory of
massive Wess-Zumino model.

\section{Conclusion}

In this paper we have investigated the applicability of the recently
developed Loop Regularization method in supersymmetric theories. By
checking several Ward identities in various supersymmetric models,
we have explicitly shown that the LR method is applicable to the
supersymmetric field theories. We have also directly carried out the
calculations for one-loop renormalization of massive Wess-Zumino
model by using the LR method with string-mode regulators, the
results are consistent with the general conclusion yielded from the
supergraph technique. Once the supersymmetric extensions of the
standard model could be discovered at the LHC, such a
symmetry-preserving Loop Regularization method with string-mode
regulators can widely be applied to the computations of various
supersymmetric processes.

\acknowledgments \label{ACK}

The authors would like to thank A.Cohen for useful discussions. This
work was supported in part by the National Science Foundation of
China (NSFC) under the grant \# 10821504, 10491306, and the Project
of Knowledge Innovation Program (PKIP) of Chinese Academy of
Science.

\appendix

\section{Translational Invariance of Loop Regularization}

The verification of translational invariance in section 2 can simply
be extended to the linearly and quadratically divergent integrals.

Consider firstly the quadratically divergent integral
\begin{equation}
L_2=\int \frac{d^4 k}{[(k-xp)^2 + M^2]}
\end{equation}
by rewriting the momentum factor $(k-xp)^2$ into $(k-xp)^2 = k^2 -
2xp.k + x^2p^2$, then replacing $k^2$ by $k^2 + M_l^2$, one has
\begin{equation}
(k-xp)^2\to k^2 + M_l^2 - 2xp.k + x^2p^2 = (k-xp)^2 + M_l^2
\end{equation}
Thus the proof in the manuscript for the scalar type logarithmic
loop integration can be easily extended to the scalar type
quadratically divergent ILIs, namely
\begin{equation}
L_2\to L_2^R = \lim_{N, M_l^2}\sum_{l=0}^{N}c_l^N\int
\frac{d^4k}{[(k-xp)^2+M_{l}^2]}
\end{equation}
The regularized ILIs $L_2^R$ is well-defined and allows us to shift
the momentum, we then have
\begin{equation}
L_2^R = \lim_{N, M_l^2}\sum_{l=0}^{N}c_l^N\int
\frac{d^4k}{[(k-xp)^2+M_{l}^2]} =\lim_{N,
M_l^2}\sum_{l=0}^{N}c_l^N\int \frac{d^4 k}{[k^2+M_{l}^2]} = I_2^R
\end{equation}

Actually, it is this translational invariance which allows us to
clarify the ambiguity caused by the linear divergent in evaluating
the triangle anomaly and CPT/Lorentz violating Chern-Simons term,
which was shown in ref. \cite{YLMa2}. To be more clear here, we
demonstrate it as follows.

Let's first present J. Jauch and F. Rohrlich's discussion on the
logarithmically divergent integrals\cite{JR}. Considering the
following integral,
\begin{equation}
L_{0}=\int \frac{d^4 k}{[(k-p)^2+M^2]^2}
\end{equation}
and making use of the identity,
\begin{equation} \label{eq:identity}
\frac{1}{\alpha^n}-\frac{1}{\beta^n}=-\int^{1}_{0}\frac{n(\alpha-\beta)dz}{[(\alpha-\beta)z+\beta]^{n+1}}
\end{equation}
for $n=2$, we can rewrite the above integral as follows
\begin{equation}\label{eq:tra}
L_{0}=\int \frac{d^4 k}{(k^2+M^2)^2}-2\int
d^4k\int^{1}_{0}\frac{(p^2-2p\cdot k)dz}{[k^2+M^2+(p^2-2p\cdot
k)z]^3} \equiv I_0 + L_c
\end{equation}
The second term $L_c$ of the right-hand side is convergent, so we
can safely shift the origin of $k$
\begin{equation}
k_{\mu}\rightarrow k_{\mu}+p_{\mu}z
\end{equation}
and the second term reads
\begin{equation}
L_c =-2\int^{1}_{0}dz\int \frac{p^2(1-2z)-2p\cdot
k}{[k^2+M^2+p^2z(1-z)]^3}d^4k
\end{equation}
The term in the numerator which is odd in k will vanish. Using the
identity
\begin{equation}
\int\frac{(k^2)^{m-2}d^4k}{(k^2+M^2)^n}=\frac{i\pi^2}{(M^2)^{n-m}}B(m,n-m)
\end{equation}
where $B(m,n-m)=\Gamma(m)\Gamma(n-m)/\Gamma(n)$ and $n>m>0$ is the
condition of convergence. So the second term in eq(\ref{eq:tra}) now
goes as
\begin{equation}
L_c =
-2\frac{i\pi^2}{2}\int^{1}_{0}dz\frac{p^2(1-2z)}{M^2+p^2z(1-z)}=-i\pi^2\ln[M^2+p^2z(1-z)]|^{1}_{0}=0
\end{equation}
Therefore, for the logarithmic divergent integral, we arrive at the
following identity
\begin{equation}
L_{0}=\int \frac{d^4 k}{[(k-p)^2+M^2]^2}=\int \frac{d^4
k}{[k^2+M^2]^2} = I_0
\end{equation}
which is independent of the regularization.

Nevertheless, if firstly applying the Loop Regularization
prescription and then shifting the momentum, the corresponding
relation becomes a straightforward consequence
\begin{equation}
L_0\to L_0^R = \lim_{N, M_l^2}\sum_{l=0}^{N}c_l^N\int
\frac{d^4k}{[(k-p)^2+M_{l}^2]^2}=\lim_{N,
M_l^2}\sum_{l=0}^{N}c_l^N\int \frac{d^4 k}{[k^2+M_{l}^2]^2} = I_0^R
\end{equation}

Let us now consider the linear divergent integral.  When using the
identity eq. (\ref{eq:identity}), a similar proof can be carried out
and shows that a shift of $k$ in a linearly divergent integral will
result in a finite additive constant
\begin{equation}
L_{1,\mu} = \int \frac{k^{\mu}d^4 k}{[(k-p)^2+M^2]^2}=\int
\frac{(k+p)^{\mu}d^4 k}{[k^2+M^2]^2}-\frac{i\pi^2}{2}p^\mu \equiv
I_{1\mu} + p_{\mu} I_0 + L_{c\mu}
\end{equation}
which has been shown to cause an ambiguity in evaluating the chiral
anomaly if the regularization schemes are not applied
appropriately\cite{YLMa2}. This is because the results may depend on
the procedure of applying the regularization schemes before or after
using the identity eq. (\ref{eq:identity}).

To be safe, we shall apply LR prescription before shifting the
momentum, it then leads to the following result
\begin{eqnarray}
L_{1,\mu}& \to & L_{1,\mu}^R = \lim_{N,
M_l^2}\sum_{l=0}^{N}c_l^N\int
\frac{k^{\mu}d^4k}{[(k-p)^2+M_{l}^2]^2} \nonumber \\
& = & \lim_{N, M_l^2}\sum_{l=0}^{N}c_l^N\int \frac{(k+p)^{\mu}d^4
k}{[k^2+M_{l}^2]^2} = \lim_{N, M_l^2}\sum_{l=0}^{N}c_l^N\int
\frac{p^{\mu}d^4 k}{[k^2+M_{l}^2]^2} = p_{\mu} I_0^R
\end{eqnarray}
where we have shifted the momentum for the well-defined regularized
integral but without using the above identity.

On the other hand, when applying LR prescription before shifting the
momentum, but using the identity presented above for the
integration, we then arrive at the following expression
\begin{eqnarray}
L_{1,\mu}\to L_{1,\mu}^R & = & \lim_{N,
M_l^2}\sum_{l=0}^{N}c_l^N\int
\frac{k^{\mu}d^4k}{[(k-p)^2+M_{l}^2]^2} \nonumber \\
& = & \lim_{N, M_l^2}\sum_{l=0}^{N}c_l^N\int \frac{(k+p)^{\mu}d^4
k}{[k^2+M_{l}^2]^2}-\frac{i\pi^2}{2}p^\mu\lim_{N,
M_l^2}\sum_{l=0}^{N}c_l^N
\end{eqnarray}
The second term of the right-hand side actually vanishes due to the
following conditions for the coefficients in LR
\begin{equation}
\lim_{N, M_l^2}\sum_{l=0}^{N}c_l^N(M_l^2)^n = 0 \quad
       (n= 0, 1, \cdots)
\end{equation}
thus we finally yield the following relation
\begin{eqnarray}
L_{1,\mu}^R & = & \lim_{N, M_l^2}\sum_{l=0}^{N}c_l^N \int
\frac{k^{\mu}d^4k}{[(k-p)^2+M_{l}^2]^2}= \lim_{N,
M_l^2}\sum_{l=0}^{N}c_l^N\int \frac{(k+p)^{\mu}d^4
k}{[k^2+M_{l}^2]^2} \nonumber \\
& = & \lim_{N, M_l^2}\sum_{l=0}^{N}c_l^N\int \frac{p^{\mu}d^4
k}{[k^2+M_{l}^2]^2} = p_{\mu} I_0^R
\end{eqnarray}
which just shows that in the LR method the translation of momentum
can safely be made for a linearly divergent integral.

Now we turn to the quadratically divergent integral,
\begin{equation}
L_{2}=\int \frac{d^4 k}{[(k-p)^2+M^2]}
\end{equation}
which can be rewritten as follows when using the previous identity
\begin{equation}\label{eq:quad}
L_{2}=\int \frac{d^4 k}{(k^2+M^2)}-2\int
d^4k\int^{1}_{0}\frac{(p^2-2p\cdot k)dz}{[k^2+M^2+(p^2-2p\cdot
k)z]^2} \equiv I_2 + L_{2c}
\end{equation}
Since the second term involves only linear and logarithmical
divergences, we can then use the previous identities for those
integrals when shifting the origin of $k$, and get the following
result with a finite additive constant
\begin{eqnarray}\label{eq:trqq}
&&L_{2c} = -2\int d^4k\int^{1}_{0}\frac{(p^2-2p\cdot
k)dz}{[k^2+M^2+(p^2-2p\cdot k)z]^2} \nonumber\\
&=&-2\int^{1}_{0}dz\int \frac{p^2d^4k}{[k^2+M^2+p^2z(1-z)]^2}\nonumber\\
&+&2\int^{1}_{0}dz\int \frac{2p\cdot
(k+xp)d^4k}{[k^2+M^2+p^2z(1-z)]^2}-i\pi^2 p^2
\end{eqnarray}
The term which is odd in $k$ does not contribute, and two integrals
of the right-hand side cancel each other due to the relation
\begin{equation}
\int^{1}_{0}dz\int
\frac{zp^2d^4k}{[k^2+M^2+p^2z(1-z)]^2}=\frac{1}{2}\int^{1}_{0}dz\int
\frac{p^2d^4k}{[k^2+M^2+p^2z(1-z)]^2}
\end{equation}
Thus we arrive at the following identity
\begin{equation}
L_2 = \int \frac{d^4 k}{[(k-p)^2+M^2]}=\int
\frac{d^4k}{[k^2+M^2]}-i\pi^2p^2 \equiv I_2 + L_{2c}
\end{equation}
Just like the discussion in linearly divergent integral, by applying
the LR prescription before shifting momentum, we have
\begin{equation}
L_2\to L_2^R = \lim_{N, M_l^2}\sum_{l=0}^{N}c_l^N\int
\frac{d^4k}{[(k-p)^2+M_{l}^2]}=\lim_{N,
M_l^2}\sum_{l=0}^{N}c_l^N\int \frac{d^4 k}{[k^2+M_{l}^2]} = I_2^R
\end{equation}
where the shift of momentum has been made for the regularized
$L_2^R$. On the other hand, again applying the LR prescript before
shifting momentum, but using the identity obtained above, we arrive
at the following expression
\begin{eqnarray}
L_2\to L_2^R & = & \lim_{N, M_l^2}\sum_{l=0}^{N}c_l^N\int
\frac{d^4k}{[(k-p)^2+M_{l}^2]} \nonumber \\
& = & \lim_{N, M_l^2}\sum_{l=0}^{N}c_l^N\int \frac{d^4
k}{[k^2+M_{l}^2]}-i\pi^2p^2\lim_{N, M_l^2}\sum_{l=0}^{N}c_l^N
\end{eqnarray}
accordingly, because of the vanish of the second term in the
right-hand side, we obtain the same regularized result
\begin{equation}
L_2^R = \lim_{N, M_l^2}\sum_{l=0}^{N}c_l^N\int
\frac{d^4k}{[(k-p)^2+M_{l}^2]}=\lim_{N,
M_l^2}\sum_{l=0}^{N}c_l^N\int \frac{d^4 k}{[k^2+M_{l}^2]} = I_2^R
\end{equation}

So far we have demonstrated that Loop Regularization can preserve
translational invariance not only in logarithmically, but also in
linearly and quadratically divergent integral.

\section{Derivation of Majorana Feynman Rules}\label{MFR}

Here we are going to present a simple and definite derivation of
Majorana Feynman rules which are useful for our calculations in this
paper. We will begin with the quantization of free Majorana fermion,
and figure out the difficulties of formulating the Majorana Feynman
rules, then provide a consistent prescription. The unusual Majorana
Feynman rules are result from the Majorana fermion self-conjugacy.
Though the two-components formulation of Majorana field is more
fundamental, it is still very useful to work in four-components
formalism because the $\gamma$ matrices is more convenient for
practical calculations.

The Majorana fermion field $\chi$ is quantized by stipulating the
following equal-time anticommutators:
\begin{eqnarray}
\{\chi_\alpha(\textbf{x}),\chi^\dag_\beta(\textbf{y})\}&=&\delta_{\alpha\beta}\delta^3(\textbf{x}-\textbf{y})\nonumber\\
\{\chi_\alpha(\textbf{x}),\chi_\beta(\textbf{y})\}&=&\{\chi^\dag_\alpha(\textbf{x}),\chi^\dag_\beta(\textbf{y})\}=0
\end{eqnarray}

The plane wave decomposition of $\chi$ is not obvious. In
two-components formalism the difficulty behaves as that the equation
of motion (EOM) is no longer a linear equation since the EOM connects
$\chi$ to its complex conjugation. In four-components formalism the
difficulty lies in the Majorana condition:
$\chi=\chi^c=C\bar{\chi}^T$. But if we use the spinors $u$ and $v$
which satisfy $u_{\textbf{k},s}=C\bar{v}^T_{\textbf{k},s}$ and
$v_{\textbf{k},s}=C\bar{u}^T_{\textbf{k},s}$, then $\chi$ can be
expanded as:
\begin{eqnarray}
\chi=\int\frac{d^3k}{(2\pi)^3}\frac{1}{2E_{\textbf{k}}}\sum_s[c_{\textbf{k},s}u_{\textbf{k},s}e^{-ikx}+c^\dag_{\textbf{k},s}v_{\textbf{k},s}e^{ikx}]\label{planewave}
\end{eqnarray}
here $c$ and $c^\dag$ are the annihilation and creation operators of
Majorana fermions. For Majorana fields, we still have:
\begin{eqnarray}
\langle 0|T\chi_\alpha(x)\bar{\chi}_\beta(y)|0\rangle=\intdk
e^{-ik(x-y)}(\frac{i}{k\sla-m})_{\alpha\beta}=S_{F\alpha\beta}(x-y)
\end{eqnarray}
Note that because of the Majorana condition $\chi=C\bar{\chi}^T$ and
$\bar{\chi}=\chi^TC$, $\langle
0|T\chi_\alpha(x)\chi_\beta(y)|0\rangle$ and $\langle
0|T\bar{\chi}_\alpha(x)\bar{\chi}_\beta(y)|0\rangle$ do not vanish.
It is easy to show that:
\begin{eqnarray}
\langle 0|T\chi_\alpha(x)\chi_\beta(y)|0\rangle&=&S_{F\alpha\gamma}(x-y)C^T_{\gamma\beta}\\
\langle
0|T\bar{\chi}_\alpha(x)\bar{\chi}_\beta(y)|0\rangle&=&C^T_{\alpha\gamma}S_{F\gamma\beta}(x-y)
\end{eqnarray}
The explicit expressions of $u_{\textbf{k},s}$ and
$v_{\textbf{k},s}$ as well as the spin-sum identities can be found
in \cite{dreiner}, we list the results here:
\begin{eqnarray}
&u_{\textbf{k},s}=\left(\begin{array}{c}
  \sqrt{k\cdot\sigma}\zeta_s\\
  \sqrt{k\cdot\bar{\sigma}}\zeta_s
\end{array}
\right),
&\bar{u}_{\textbf{k},s}=\left(\zeta^\dag_s\sqrt{k\cdot\bar{\sigma}},\zeta^\dag_s\sqrt{k\cdot\sigma}\right)\nonumber\\
&v_{\textbf{k},s}=\left(\begin{array}{c}
  2s\sqrt{k\cdot\sigma}\zeta_{-s}\\
  -2s\sqrt{k\cdot\bar{\sigma}}\zeta_{-s}
\end{array}\right),
&\bar{v}_{\textbf{k},s}=\left(-2s\zeta^\dag_s\sqrt{k\cdot\bar{\sigma}},2s\zeta^\dag_s\sqrt{k\cdot\sigma}\right)
\end{eqnarray}
and $\zeta_{\pm1/2}$ are defined as below (here $\theta$ is the polar angle of $\textbf{k}$, and $\phi$ is the azimuthal angle of $\textbf{k}$.):
\begin{eqnarray}
\zeta_{1/2}(\textbf{k})=\left(\begin{array}{c}
  cos\frac{\theta}{2}\\
  e^{i\phi}sin\frac{\theta}{2}
\end{array}\right),
\hspace{1cm}
\zeta_{-1/2}(\textbf{k})=\left(\begin{array}{c}
  -e^{-i\phi}sin\frac{\theta}{2}\\
  cos\frac{\theta}{2}
\end{array}\right)
\end{eqnarray}
The spin-sum identities are:
\begin{eqnarray}
\sum_s u_{\textbf{k},s}\bar{u}_{\textbf{k},s}&=&k\sla+m \nonumber \\
\sum_s v_{\textbf{k},s}\bar{v}_{\textbf{k},s}&=&k\sla-m \nonumber \\
\sum_s u_{\textbf{k},s}v^T_{\textbf{k},s}&=&(k\sla+m)C^T \nonumber \\
\sum_s v_{\textbf{k},s}u^T_{\textbf{k},s}&=&(k\sla-m)C^T\\
\sum_s \bar{u}^T_{\textbf{k},s}\bar{v}_{\textbf{k},s}&=&C^\dag(k\sla-m) \nonumber \\
\sum_s
\bar{v}^T_{\textbf{k},s}\bar{u}_{\textbf{k},s}&=&C^\dag(k\sla+m)
\nonumber
\end{eqnarray}

Before starting the derivation of Majorana Feynman rules we may briefly review
the derivation for the usual Dirac fermions. The argument below follows the one in
\cite{denner}. The calculation of a typical scattering matrix
element corresponds to the evaluation of the following expression:
\begin{eqnarray}
\langle
0|b_1...b_md_1...d_nT[(\bar{\psi}(x_1)\Gamma\psi(x_1))...(\bar{\psi}(x_l)\Gamma\psi(x_l))]b^\dag_1...b^\dag_pd^\dag_1...d^\dag_q|0\rangle
\end{eqnarray}
Firstly, we should rearrange the interaction terms to make them
following the order of contractions. Since only one type of
contraction $\langle\psi\bar{\psi}\rangle$ exists for Dirac fermion,
the internal propagator reads:
$\langle\psi\bar{\psi}\rangle=S_F(p)$, here the fermion charge and
the momentum flows are well defined from $\bar{\psi}$ to $\psi$, the
Feynman rule for vertex directly reads as $i\Gamma$. For Dirac
fermion, the fermion charge flow (in fact this is also the momentum
flow) of internal popagator forms a continuous flow, when writing
down the analytic expression one should first do it oppositing to
the continuous flow. The most important step is to determine the
Relative Sign of Interfering Feynman graphs (RSIF). There are in
general three types of commutations which can contribute to the
RSIF. Firstly, when reordering $b_i$, $d_i$, $b^\dag_i$ and
$d^\dag_i$ to put them in the appropriate places of Wick
contractions, it causes a factor $(-1)^P$. Here $P$ is the parity of
the permutation of the annihilation and creation operators. This
factor can be read from the order of external spinors in the
analytic expression with respect to the given reference order.
Secondly, for a closed fermion loop, one needs to exchange the first
and the last field operator in the fermion chain, which gives a
factor $(-1)^L$, where $L$ is the number of fermion loops. Finally,
since $d^\dag_i$ must contract with $\bar{\psi}$ and $d_i$ must
contract with $\psi$, one needs to move the creation operator $d_i$
to the beginning of Wick contraction and move the annihilation
operator $d^\dag_i$ to the end, which leads to a factor $(-1)^V$
with $V$ being the total number of spinors $v$ and $\bar{v}$. Since
$V$ is universal for all graphs of a given process, this factor can
therefore be omitted.

We now trun to investigate the Majorana fermion case. Firstly, we
consider the situation that there are no Dirac fermions but only
Majorana fermions. As mentioned above, all possible contractions
between $\chi$ and/or $\bar{\chi}$ do not vanish now. In this case,
after rearranging the interaction terms to perform Wick contraction
for operators one by one, we need to consider four types of Majorana
propagators, i.e. $\langle\chi\chi\rangle$,
$\langle\chi\bar{\chi}\rangle$, $\langle\bar{\chi}\chi\rangle$ and
$\langle\bar{\chi}\bar{\chi}\rangle$. More seriously, the
propagators depend on the sign of its momentum $p$, but now we can
not define the orientation from $\bar{\chi}$ to $\chi$ as the arrow
of momentum. That means we need to find out a new method to resign
the arrow of momentum. For the Feynman rule of vertex, it raises a
new ambiguity. For instance, when contracting an interaction
Lagrangian $\bar{\chi}\Gamma\chi$ in the time-order product, one can
contract the operator $\bar{\chi}$ with one field operator lies on
the left of this vertex and contract $\chi$ with another lies on the
right, or one can also contract $\chi$ with one field operator lies
on the left and contract $\bar{\chi}$ with another lies on the
right. In the later case an additional $(-1)$ will emerge. Previous
discussions\cite{gates, gluza} for the Majorana Feynman rules
follows this analysis and try to reduce the number of propagators
and vertices, while the resulting consequences are still too obscure
and not easy to use. In ref\cite{denner}, the author introduced the
charge-conjugate fields $\psi^c$ and $\bar{\psi^c}$ to Feynman rules
and tried to give a uniform description of Dirac and Majorana field.
Here we shall provide an alternative and simple description.

Firstly, we may eliminate $\bar{\chi}$ from the interaction
Lagrangian by using the Majorana condition
$\bar{\chi}=-\chi^TC^\dag$, so that only one type of propagator
$\langle\chi\chi\rangle$ remains. We then use a line without arrow
to represent a Majorana propagator. Since Majorana fermions can not
carry any charge, this representation is natural. In the momentum
space, the Feynman rule for Majorana propagator is
$\frac{i}{k\sla-m}C^T$. To obtain the Feynman rule of vertex, we may
rewrite $\bar{\chi}\Gamma\chi$ as:
\begin{eqnarray}
\bar{\chi}_\alpha\Gamma_{\alpha\beta}\chi_\beta&=&\chi_\alpha(-C^\dag_{\alpha\rho}\Gamma_{\rho\beta})\chi_\beta=-\chi_\beta(\Gamma^T_{\beta\rho}C^\dag_{\rho\alpha})\chi_\alpha=\frac{1}{2}\chi_\alpha(-C^\dag\Gamma-\Gamma^TC^\dag)_{\alpha\beta}\chi_\beta\nonumber\\
&=&\frac{1}{2}\chi_\alpha\Gamma^\prime_{\alpha\beta}\chi_\beta
\end{eqnarray}
with:
\begin{eqnarray}
\Gamma^\prime=-C^\dag\Gamma-\Gamma^TC^\dag=-\Gamma^{\prime T}
\end{eqnarray}
Now the ambiguity mentioned about disappears as
$\Gamma^\prime$ is antisymmetric. The Feynman rule for vertex simply
becomes: $i\Gamma^\prime$. One can treat the Majorana fermions just
like a real scalar boson to obtain the correct symmetric factor of a
given graph.

Next, we should determine the direction of momentum in Majorana
propagators. Remember that generally a factor $e^{-ikx}$ means
momentum $k$ flows in the point $x$ and $e^{ikx}$ means momentum k flows
out the point $x$. Every contraction between two field operators $O(x)$,
$O(y)$ can always be written in the form: $\langle
O(x)O(y)\rangle=\intdk e^{-ik(x-y)}S(k)$, for example in our case:
\begin{eqnarray}
\langle 0|T\chi_\alpha(x)\chi_\beta(y)|0\rangle=\intdk
e^{-ik(x-y)}(\frac{i}{k\sla-m}C^T)_{\alpha\beta}
\end{eqnarray}
which indicates that the direction of momentum flow is always
opposite to the direction of contraction for a propagator, and in a
fermion chain the momentum flows of propagators form a continuous
flow its direction is opposite to the direction of contractions. In
\cite{denner} such a folw was called as 'fermion flow', here we may,
more precisely, call it as 'fermion momentum flow'. This comes to
the conclusion: for each fermion chain we fix an arbitrary
orientation (fermion momentum flow), the momentums of all fermion
propagators follow this orientation, and we should write down the
Feynman rules proceeding opposite to the chosen orientation.

Finally, to complete the Majorana Feynman rules, it needs to give
the rules of external fermion lines and determine the RSIF. The
rules of external fermion lines can easily be obtained from the
plan-wave decomposition of $\chi$, see Eqn.(\ref{planewave}). Since
\begin{eqnarray}
\langle 0|c_{\textbf{k},s}\chi_\alpha(x)&\longrightarrow&v_{\alpha\textbf{k},s}e^{ikx}\\
\chi_\alpha(x)c^\dag_{\textbf{k},s}|0\rangle&\longrightarrow&u_{\alpha\textbf{k},s}e^{-ikx}
\end{eqnarray}
which implies that the creation of a Majorana fermion corresponds to
a spinor $v_{\alpha\textbf{k},s}$ with momentum $k$ flow out, and
the annihilation of a Majorana fermion corresponds to a spinor
$u_{\alpha\textbf{k},s}$ with momentum $k$ flow in. If the spinor
locates at the beginning of contraction, we should write it as a row
vector say a $u^T$ or $v^T$. Now we can give a prescription to fix
RSIF. Factor $(-1)^P$ can be got from the permutation parity of the
spinors in the obtained analytical expression with respect to some
reference order. Factor $(-1)^L$ can be got from the number of
closed fermion loops. Factor $(-1)^V$ now is a little different from
which in Dirac field theory. Since moving any one creation operator
arising from the initial state to the beginning of contraction will
contribute a factor $-1$, and moving any one annihilation operator
arising from the final state to the end of contraction which also
contribute a factor $-1$, it seems that we should count the total
number of such operation. Suppose that there are '$a$' fermions in
the initial sate and '$b$' fermions in the final state, and we must
move $i$th fermion creation operators to the beginning and $j$th
fermion annihilation operators to the end, then we have
$a-i+j=b+i-j$, i.e. $|i-j|=\frac{1}{2}|a-b|$. Namely,
$V=\frac{1}{2}|a-b|$. Since $a$ and $b$ is universal for all graphs
of a process, we can always omit $(-1)^V$ all the time.

Let us consider the situation that a Majorana fermion $\chi$ couples
to a Dirac fermion $\psi$, the interaction Lagrangian contains the
following terms:
\begin{eqnarray}
\bar{\chi}\Gamma\psi+\bar{\psi}\bar{\Gamma}\chi \hspace{1cm}(where:
\bar{\Gamma}=\gamma^0\Gamma^\dag\gamma^0)
\end{eqnarray}

When keeping a continuous "fermion momentum flow" for a fermion
internal line, we then need to consider two types of Dirac
propagators: $\langle\psi\bar{\psi}\rangle$ and
$\langle\bar{\psi}\psi\rangle$ which have the following explicit
forms:
\begin{eqnarray}
\langle 0|T\psi_\alpha(x)\bar{\psi}_\beta(y)|0\rangle&=&\intdk e^{-ik(x-y)}(\frac{i}{k\sla-m})_{\alpha\beta}\\
\langle 0|T\bar{\psi}_\alpha(x)\psi_\beta(y)|0\rangle&=&\intdk
e^{-ik(x-y)}[(\frac{i}{k\sla+m})^T]_{\alpha\beta}
\end{eqnarray}
We then need to use a line with arrow to represent the Dirac
propagator, the arrow reflects the flow of charge which flows out of
$\bar{\psi}$ and into $\psi$. If the direction of charge flow
coincide with the direction of the "fermion momentum flow", we
should use $\langle\psi\bar{\psi}\rangle=\frac{i}{k\sla-m}$,
otherwise we should use
$\langle\bar{\psi}\psi\rangle=(\frac{i}{k\sla+m})^T$.

The Feynman rules for vertexes are also doubled. For the Dirac-Dirac
interaction, one has:
\begin{eqnarray}
\bar{\psi}_\alpha\Gamma_{\alpha\beta}\psi_\beta=\psi_\alpha(-\Gamma^T)_{\alpha\beta}\bar{\psi}_\beta
\end{eqnarray}
If the direction of charge flow coincide with the direction of the
momentum flow, we should use $i\Gamma$, otherwise we should use
$-i\Gamma^T$. the vertexes rules of Majorana-Dirac interaction can
be derived similarly from the identities:
\begin{eqnarray}
\bar{\chi}_\alpha\Gamma_{\alpha\beta}\psi_\beta&=&\chi_\alpha(-C^\dag\Gamma)_{\alpha\beta}\psi_\beta=\psi_\alpha(-\Gamma^TC^\dag)_{\alpha\beta}\chi_\beta\\
\bar{\psi}_\alpha\bar{\Gamma}_{\alpha\beta}\chi_\beta&=&\chi_\alpha(-\bar{\Gamma}^T)_{\alpha\beta}\bar{\psi}_\beta
\end{eqnarray}

The RSIF can be determined by using the same method as we mentioned
above.

With the above considerations, we can summarize our Feynman rules.
The solid lines are still used to denote the fermions. Dirac
fermions lines carry arrows which reflect the direction of charge
flow, Majorana lines do not carry arrows. We may write down Feynman
amplitudes according to the following steps:

1. Draw all topologically distinctive, connect Feynman diagrams for
a given process.

2. Fix an arbitrary direction for each fermion chain. This is the
direction of "fermion momentum flow", which means that the momentum
of every internal fermion line should follow this direction. We
should write down the Dirac matrices proceeding opposite to the
chosen direction through the chain.

\begin{figure}[ht]
\includegraphics[scale=0.7]{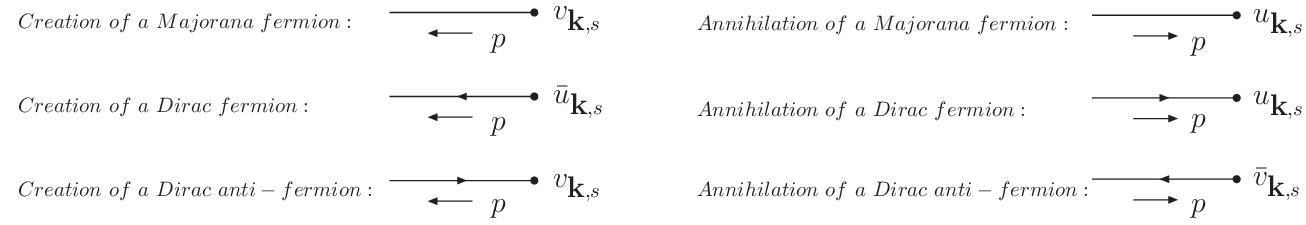}
\caption{Feynman rules for external lines}\label{external}
\end{figure}

3. For the external fermion lines, the rules are shown in
FIG.\ref{external}.

If the spinors locates at the beginning(end) of contraction, we
should add a superscript $T$ appropriately to write them as
row(column) vectors.

\begin{figure}[ht]
\includegraphics[scale=0.7]{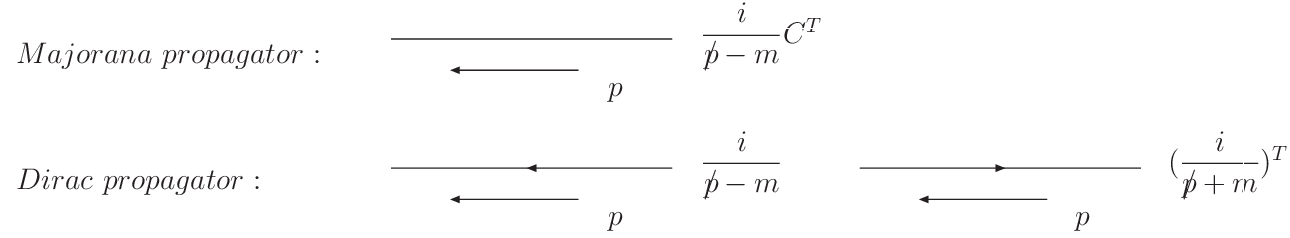}
\caption{Feynman rules for propagators}\label{propagator}
\end{figure}

4. For the fermion propagators, the rules are shown in
FIG.\ref{propagator}.

\begin{figure}[ht]
\includegraphics[scale=0.7]{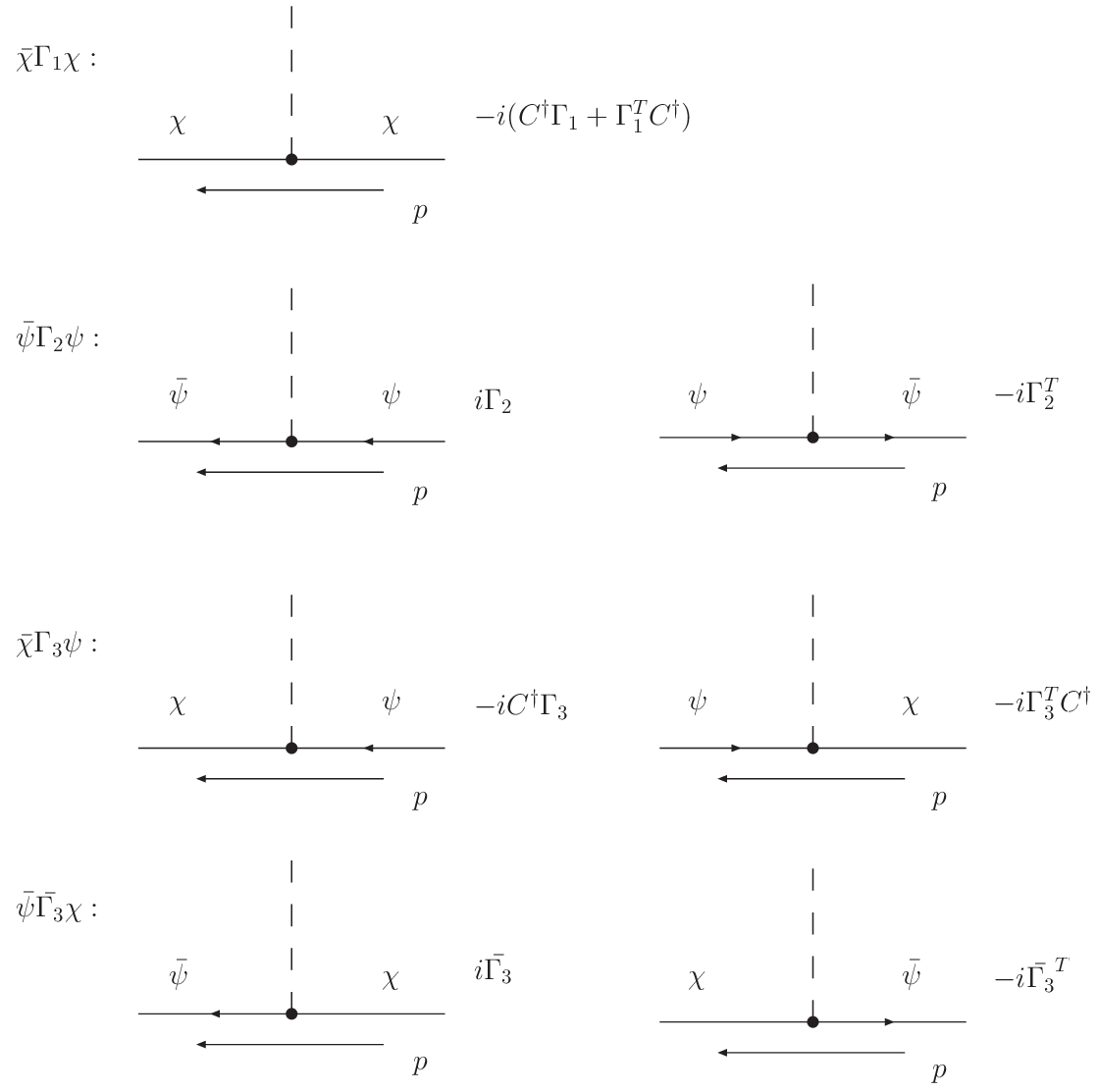}
\caption{Feynman rules for vertexes}\label{vertex}
\end{figure}

5. For the general fermion interactions $\bar{\chi}\Gamma_1\chi$,
$\bar{\psi}\Gamma_2\psi$,
$\bar{\chi}\Gamma_3\psi+\bar{\psi}\bar{\Gamma_3}\chi$, where
$\bar{\Gamma_3}=\gamma^0\Gamma_3^\dag\gamma^0$, the Feynman
rules are shown in FIG.\ref{vertex} respectively.

6. To determine the RSIF. For each diagram, multipling by a
factor $(-1)$ for each closed fermion loop, and multipling by the
permutation parity of the spinors in the obtained analytical
expression with respect to some reference order.

7. Multipling a symmetry factor $S^{-1}$ for each diagram. The Majorana fermions may be treated just as real scalar fields to obtain the
symmetry factor.
\begin{eqnarray}
S=g\prod_{n=2,3,...}2^\beta(n!)^{\alpha_n}
\end{eqnarray}
where $\alpha_n$ is the number of pairs of vertices connected by $n$
identical self-conjugate lines, $\beta$ is the number of lines
connecting a vertex with itself, and $g$ is the number of
permutations of vertices which leave the diagram unchanged with
fixed external lines.

\newpage

\section{Feynman Rules of Massive Wess-Zumino model}\label{FRoWZ}
\begin{figure}[ht]
\includegraphics[scale=0.7]{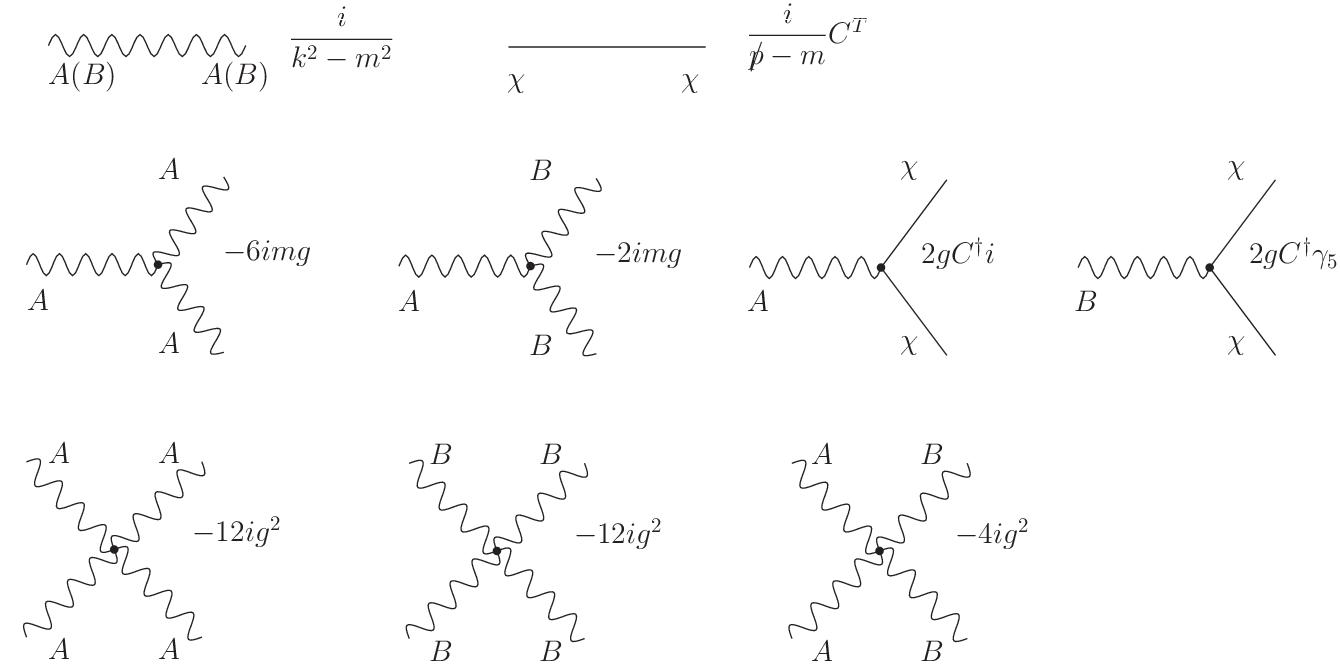}
\caption{Feynman rules of massive Wess-Zumino model}\label{wzfrules}
\end{figure}

We present all the Feynman rules of this model in FIG.\ref{wzfrules}.

\newpage


\end{document}